\documentclass[5p,times]{elsarticle}

\usepackage{amssymb}
\usepackage{amsmath}
\usepackage{acronym}
\usepackage{caption}
\usepackage{acronym}
\usepackage{multirow}
\usepackage{lscape}
\usepackage{booktabs}
\usepackage{longtable}
\usepackage{url}
\usepackage{wrapfig}
\usepackage[section]{placeins}

\newacro{SKA}{Square Kilometre Array}
\newacro{MWA}{Murchison Widefield Array}
\newacro{FFT}{Fast Fourier Transform}
\newacro{UML}{Unified Modeling Language}
\newacro{LGT}{Logical Graph Template}
\newacro{LG}{Logical Graph}
\newacro{PGT}{Physical Graph Template}
\newacro{PG}{Physical Graph}
\newacro{RG}{Runtime Graph}
\newacro{DAG}{Directed Acyclic Graph}
\newacro{DALiuGE}{Data Activated Flow Graph Engine}
\newacro{BlockDAG}{block directed acyclic graph}
\newacro{NCC}{normalised cross correlation}

\journal{Elsevier}

\begin{document}

\begin{frontmatter}



\title{Formal Definition and Implementation of Reproducibility Tenets for Computational Workflows}

\affiliation[icrar]{organization={International Centre for Radio Astronomy Research, University of Western Australia},
             addressline={7 Fairway},
             city={Perth},
             postcode={6009},
             state={WA},
             country={Australia}}

\author[icrar]{Nicholas J. Pritchard} 
\ead{nicholas.pritchard@icrar.org}
\author[icrar]{Andreas Wicenec}
\ead{andreas.wicenec@icrar.org}
\begin{abstract}
Computational workflow management systems power contemporary data-intensive sciences.
The slowly resolving reproducibility crisis presents both a sobering warning and an opportunity to iterate on what science and data processing entails.
The \ac{SKA}, the world's largest radio telescope, is among the most extensive scientific projects underway and presents grand scientific collaboration and data-processing challenges.
In this work, we aim to improve the ability of workflow management systems to facilitate reproducible, high-quality science.
This work presents a scale and system-agnostic computational workflow model and extends five well-known reproducibility concepts into seven well-defined tenets for this workflow model.
Additionally, we present a method to construct workflow execution signatures using cryptographic primitives in amortized constant time.
We combine these three concepts and provide a concrete implementation in \ac{DALiuGE}, a workflow management system for the \ac{SKA} to embed specific provenance information into workflow signatures, demonstrating the possibility of facilitating automatic formal verification of scientific quality in amortized constant time.
We validate our approach with a simple yet representative astronomical processing task: filtering a noisy signal with a lowpass filter using CPU and GPU methods. This example shows the practicality and efficacy of combining formal tenet definitions with a workflow signature generation mechanism.
Our framework, spanning formal UML specification, principled provenance information collection based on reproducibility tenets, and finally, a concrete example implementation in \ac{DALiuGE} illuminates otherwise obscure scientific discrepancies and similarities between principally identical workflow executions.
\end{abstract}



\end{frontmatter}

\section{Introduction}\label{sec:intro}
Reuse is the ultimate goal of science; by building trust in observations, theories, models, and experiences, humanity explores the universe.
The ability to reproduce elements of scientific works increases our trust in their findings.
The reproducibility crisis \citep{baker_1500_2016} heightened scrutiny and discussion around what highly trusted modern science entails.
The \ac{SKA} is an already storied behemoth of academic collaboration that demands paying attention to scientific reproducibility.
Radio astronomy relies heavily on computational methods but is fundamentally an experimental, observational science, and hence, it presents a curious case for scientific reproducibility.
Seeking a balance between the world of formal software engineering and the exploratory nature of scientific investigation has already proved challenging to achieve \citep{peng_reproducibility_2015, perkel_challenge_2020}.
The \ac{SKA}, representing an order of magnitude increase in data-processing requirements on current observatories \citep{quinn_delivering_2015}, benefits from a focus on building scientific reproducibility into its instrumentation, providing an opportunity to improve what reproducibility practically means for astronomy and other computationally powered sciences.
The literature widely debates the precise meaning of scientific reproducibility and replicability \citep{barba_terminologies_2018, gundersen_fundamental_2020} and, concerning computing, focuses on `forwards facing' approaches to guarantee the future reproducibility of current computations.
While practically valuable, the general trend toward adhering to specific technologies and conventions is potentially ruinous from a philosophical standpoint; how can we know that our efforts to make scientific works reproducible succeed?
This work presents and validates a scale and system agnostic framework for `backwards facing' reproducibility tests on computational workflow executions, facilitating efficient verification of scientific reproducibility for astronomical workflows.
Specifically, we make the following contributions:
\begin{enumerate}
    \item Present a scale-agnostic computational workflow UML model suitable for computational reproducibility discussions.
    \item Define seven reproducibility tests that we term tenets, based on known literature.
    \item Describe a hash-graph-based workflow signature method, termed BlockDAGs, based on the scale-agnostic model allowing for constant time generation of workflow signatures.
    \item Implement this signature mechanism in the \ac{DALiuGE} \citep{wu_daliuge:_2017} workflow management system combined with provenance information collection guided by the workflow model and tenet definitions.
    \item Demonstrate the efficacy of the workflow model, signature method, and reproducibility tenet definitions with a demonstrative lowpass filter workflow.
\end{enumerate}
While the reproducibility tenets we present have appeared in related contexts before, the main contribution of our work lies in integrating these tenets together into a novel testing mechanism. This mechanism conceptually and technically synthesizes elements from across the literature, offering a unified approach that advances beyond a simple presentation of existing tenets or by taking a singular view of computational workflow reproducibility.

\section{Related Work}
Reproducibility is fundamental to science, and computational reproducibility in support of quality science is a long-standing goal in the field, but one where the very concept is confused.
Computational workflow managers are among the most powerful tools available in this quest, which, alongside contemporary software engineering techniques and provenance capture, are moving towards a world of fully Rerunnable, Repeatable, Reproducible, and Replicable research.
Claerbout \cite{claerbout_electronic_2005} coined the concept of computational reproducibility in scientific computing, and Peng \cite{peng_reproducible_2011} elaborates on what is needed to achieve computational reproducibility in science.
Drummond \cite{drummond_replicability_2009} separates the concepts of reproducibility and applicability.
Barba \cite{barba_terminologies_2018} reviews the myriad of terminologies used to describe computational reproducibility as Benureau and Rougier \cite{benureau_re-run_2018} provide a framework separation Rerunning, Repeating, Reproducing, and Replicating computational works.
Central to these discussions is a separation between bitwise precise replication of exact results and conceptual reproductions of an existing work independent of precise implementation.
Our definitions of these Reproducibility tenets unify these two views.
Current suggested best practices in pursuit of scientific reproducibility include simple rules \cite{sandve_ten_2013}, adapting modern tooling from industry including notebook environments \cite{ram_git_2013, boettiger_introduction_2015, beg_using_2021}, sharing code \cite{nemiroff_astrophysics_1999} and publishing openly \cite{ginsparg_about_1991, smith_journal_2018}.
Related to the signature approach we present in this work, Bellini \cite{bellini_blockchain_2019} investigates the feasibility of building a DOI-like dataset identifier leveraging public blockchains.

Computational reproducibility is a driving motivation in creating scientific workflow management systems.
In the world of computational workflow management systems, reproducibility often relates to storing workflow descriptions as scientific artifacts themselves \cite{lamprecht_towards_2020, goble_fair_2020}, according to specific standards such as FAIR \cite{wilkinson_fair_2016} for later investigation and use.
da Silva et al. \cite{ferreira_da_silva_characterization_2017} cite dynamic provenance capture as necessary for workflows to scale larger and tackle more challenging scientific problems. Gaignard et al. \cite{gaignard_domain-specific_2014} leverage semantic web technologies to produce valuable workflow provenance information, later expanding this concept into a workable ontology \cite{gaignard_findable_2020}.

Highly scalable and trustable computing is critical to conquering our largest scientific endeavors \cite{schaduangrat_towards_2020}, and as we move towards reproducible computational workflow management, we must also ensure these efforts function as expected; a key motivation behind our signature-based approach to workflow reproducibility.

\section{Computational Workflows}\label{sec:compworkflows}
Computational workflows (herein, `workflows') are structured assemblies of computational tasks and data stores \citep{liew_scientific_2017} and underpin some of the world's most extensive science projects, including the \ac{SKA} \citep{quinn_delivering_2015}.
Workflows separate domain-specific software knowledge from scientific methodology, making increasingly complex computations possible for increasingly more scientists, and have seen broad uptake in the past three decades \citep{ferreira_da_silva_workflows_2021}.
Many workflow management systems provide several flavors of workflows for specific science domains \citep{atkinson_scientific_2017}, and thus, any attempt to address workflow reproducibility should be tool-agnostic.
This section introduces a generic computational workflow model and discusses potential mappings to several well-known workflow management systems before delving into a concrete mapping for \ac{DALiuGE} \citep{wu_daliuge:_2017}, \citep{wang_summit:2019}, a workflow management system developed to execute workflows at the scale the \ac{SKA} requires.
Providing a general workflow model allows reasoning about computational reproducibility applied to the model to apply to other mapped workflow systems. Such interoperability of reasoning is necessary in working towards a holistic approach to computational reproducibility in data-intensive sciences. 

\subsection{Scale and System Agnostic Workflow Model}\label{sec:meth:model}
\begin{figure*}[htbp]
	\centering
	\includegraphics[width=0.5\textwidth]{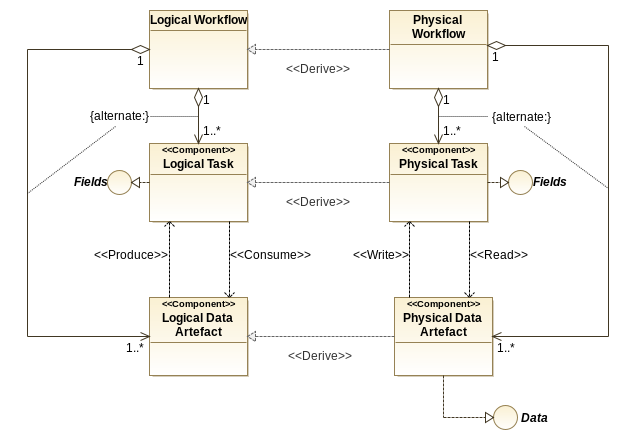}
    \caption{A \ac{UML} diagram depicting an arbitrary-scale workflow definition.\\
	\textbf{Scientific information} effectively expresses the information used in making a scientific claim and is derived by running a workflow.\\
	A \textbf{workflow} is a collection of components and data artifacts with an imposed ordered structure.\\
	A \textbf{logical workflow} is a workflow comprised of logical components.\\
	A \textbf{physical workflow} is a workflow comprised of physical components.\\
	A \textbf{component} is an atomic digital entity; it could be a (logical or physical) task or data artifact.\\
	A \textbf{logical task} describes a task characterized by high-level information like programming language, algorithm, and required options and parameters. \\
    \textbf{Logical task fields} capture additional detail about a logical task, such as the exact software script, package, or command executed. \\
	A \textbf{physical task} is a single, executable instance of a logical task. A logical task could be realized by one or more physical tasks depending on the degree of parallelism of the workflow. \\
    \textbf{Physical task fields} capture additional detail about a physical task's execution, such as the machine details a task is eventually executed on. \\
	A \textbf{logical data artifact} is a logical data resource characterized by storage type and required configurations.\\
	A \textbf{physical data artifact} is the actual datastore like a file, distributed file system, or database. A single logical data artifact may encompass many individual physical data artifacts at runtime, dependent on the degree of parallelism of the workflow.\\\\
	This ordered structure requires components and data artifacts to appear in alternating order; another component cannot precede a component.}
	\label{fig:WorkflowModel}
\end{figure*}

Figure \ref{fig:WorkflowModel} presents a generalized \ac{UML} workflow model. We intend for this model to make reasoning about testing workflow executions concrete and applicable to arbitrary workflow management systems.
Introducing a universal representation makes discussing, understanding, and implementing subsequently described reproducibility tests easier.
A \ac{UML} definition makes all modeled workflows well-defined software in their own right and permits native workflow implementation independent of using a workflow management system.
Moreover, we separate logical and physical workflow components.
This separation captures the intuitive experience of designing or describing workflows as a separate entity before executing a workflow and allows for reasoning and comparison of logical workflow designs irrespective of their eventual implementation in software as physical components.
Moreover, several physical components can, in principle, implement a single logical workflow component, as is the case in large-scale workflows. 
Finally, the model's requirement to alternate computing tasks and data artifacts ensures that a workflow description captures all interim and transient data artifacts.
This lightweight model aims to flexibly model most existing workflow management system behaviors, capturing the difference between designing and executing workflows and between elements that compute and those that store data.
While further model formalization, including ontologies for generalized scientific processes, allows accurate descriptions of data provenance \citep{celebi_towards_2020}, further formalization is beyond the scope of this research, whose focus lies on direct software implementation and testing of high-level reproducibility concerns.
\subsubsection{Mapping the workflow model to workflow management systems}
\begin{table*}[!htbp]
\centering
\caption{Mapping from workflow model to workflow management system components. Although this work presents only an implementation for DALiuGE, we intend that the reasoning applied to the workflow model can be applied to other workflow management systems.}
\label{tab:workflowmappings}
\resizebox{\textwidth}{!}{%
\begin{tabular}{@{}cccccccc@{}}
\toprule
\begin{tabular}[c]{@{}c@{}}Workflow\\ Model\\ Element\end{tabular} & DALiuGE                                                          & Taverna                                                             & Askalon                                                       & Pegasus                                                             & Kepler         & Nextflow                                                      & REANA                                                           \\ \midrule
Component                                                          & Component                                                        & -                                                                   & Action                                                        & -                                                                   & Component      & -                                                             & -                                                               \\ \midrule
\begin{tabular}[c]{@{}c@{}}Logical\\ Task\end{tabular}             & \begin{tabular}[c]{@{}c@{}}Logical\\ Component\end{tabular}      & Activity                                                            & \begin{tabular}[c]{@{}c@{}}Activity\\ Type\end{tabular}       & \begin{tabular}[c]{@{}c@{}}Abstract\\ Task Nodes\end{tabular}       & Actor/Director & Process                                                       & \begin{tabular}[c]{@{}c@{}}Container\\ Description\end{tabular} \\ \midrule
\begin{tabular}[c]{@{}c@{}}Physical\\ Task\end{tabular}            & \begin{tabular}[c]{@{}c@{}}Application\\ Drop\end{tabular}       & Processor                                                           & \begin{tabular}[c]{@{}c@{}}Activity\\ Deployment\end{tabular} & \begin{tabular}[c]{@{}c@{}}Executable\\ Task Nodes\end{tabular}     & Executable     & Process                                                       & \begin{tabular}[c]{@{}c@{}}Container\\ Instance\end{tabular}    \\ \midrule
\begin{tabular}[c]{@{}c@{}}Logical\\ Data Artifact\end{tabular}    & \begin{tabular}[c]{@{}c@{}}Logical\\ Data Component\end{tabular} & Data Link                                                           & Data Ports                                                    & \begin{tabular}[c]{@{}c@{}}Edges /\\ Logical Filenames\end{tabular} & Data Ports     & Filename                                                      & Filename                                                        \\ \midrule
\begin{tabular}[c]{@{}c@{}}Physical\\ Data Artifact\end{tabular}   & Data Drop                                                        & Data Item                                                           & Systems Files                                                 & System Files                                                        & System Files   & \begin{tabular}[c]{@{}c@{}}File /\\ Input Stream\end{tabular} & File Contents                                                   \\ \midrule
\begin{tabular}[c]{@{}c@{}}Logical\\ Workflow\end{tabular}         & Logical Graph                                                    & \begin{tabular}[c]{@{}c@{}}Workflow\\ Object (model)\end{tabular}   & \begin{tabular}[c]{@{}c@{}}Compound\\ Activity\end{tabular}   & \begin{tabular}[c]{@{}c@{}}Abstract\\ Workflow\end{tabular}         & Model          & Pipeline                                                      & \begin{tabular}[c]{@{}c@{}}Workflow\\ Description\end{tabular}  \\ \midrule
\begin{tabular}[c]{@{}c@{}}Physical\\ Workflow\end{tabular}        & Physical Graph                                                   & \begin{tabular}[c]{@{}c@{}}Executable\\ Workflow Graph\end{tabular} & Activity Set                                                  & \begin{tabular}[c]{@{}c@{}}Executable\\ Workflow\end{tabular}       & Workflow       & Workflow                                                      & Workflow                                                        \\ \bottomrule
\end{tabular}%
}
\end{table*}
Mapping between specific workflow management systems and a generic workflow model facilitates a consistent approach to reasoning about reproducibility.
While reasoning about any possible workflow management systems is likely impossible, Table \ref{tab:workflowmappings} presents possible mappings for several well-known workflow management systems.
Aiming to provide a good cross-section of approaches and use cases for workflow management systems, the review da Silva et al. provide \cite{ferreira_da_silva_characterization_2017} motivates our selection.
For each workflow management system presented, we provide a brief description of its history, usage, and reproducibility efforts concerning that particular system.

\emph{Taverna} was among the first scientific workflow management systems designed primarily for orchestrating web services as data-processing workflows \cite{ferreira_da_silva_characterization_2017}. Its web-service-oriented approach was widely adopted in bioinformatics and influenced subsequent workflow management systems. Taverna's ability to package workflows as separate research objects sets a precedent for workflow sharing and reuse. In mapping to the generic workflow model, Taverna workflows are expressed as graphs of computational components, typically web services, with a distinction between logical and physical data captured through a data manager \cite{missier_taverna_2010}. Taverna's early adoption of workflow provenance remains a core feature \cite{oinn_taverna_2004, wolstencroft_taverna_2013}, though it does not explicitly focus on reproducibility and reusability.

\emph{Askalon} is a pioneering workflow management system focused on grid computing, utilizing AGWL, an XML schema for defining workflows \cite{fahringer_askalon_2005}. Askalon separates workflow design from execution concerns, allowing novel scheduling and runtime environments. Its robust schema-based approach makes it an ideal candidate for mapping to a generic workflow model, particularly in expressing parallelism and scalability. Mapping Askalon to the model is straightforward due to its structured approach, though the lack of explicit logical data objects in place of data ports poses challenges. Despite its pre-reproducibility crisis design, Askalon’s rigorous workflow schema guarantees a level of correctness and uniqueness.

\emph{Pegasus} is a workflow management system designed for executing workflows in distributed environments, with a strong focus on scalability and reliability \cite{deelman_pegasus_2005, deelman_pegasus_2015}. It has been widely used across various scientific domains, making it a canonical choice for analysis \cite{ferreira_da_silva_characterization_2017}. Pegasus workflows are expressed as directed acyclic graphs (DAGs) in XML, which aligns well with the generic workflow model's structure. This alignment facilitates the mapping process, especially in distinguishing between logical and physical tasks. Pegasus's intelligent features, such as the replica catalog \cite{deelman_pegasus_2015}, enhance performance and reduce recomputation, though the dynamic nature of distributed environments poses challenges for reproducibility.

\emph{Kepler} is a scientific workflow management system that leverages a unique actor-director model inherited from the Ptolemy II Java library to support a wide range of scientific domains \cite{ludascher_scientific_2006, ferreira_da_silva_workflows_2021}. The flexibility of this model allows Kepler to adapt workflows without altering their logic, making it a significant system for mapping to the generic workflow model. Kepler's approach to workflow enactment, coupled with its forward-thinking data provenance system \cite{altintas_provenance_2006, missier_w3c_2013, moreau_open_2011}, ensures that workflows can be partially Rerun, enhancing reproducibility. However, the actor-director paradigm introduces complexity in mapping logical tasks to physical executions within the generic workflow model.

\emph{Nextflow} is a modern workflow management system built from open-source components, particularly popular in bioinformatics for its performance and a wide array of supported executors \cite{di_tommaso_nextflow_2017}. It uses a domain-specific language to assemble workflows from processes and files, which aligns well with the generic workflow model. Nextflow’s emphasis on containerization and reproducibility makes it straightforward to map its elements to the model, ensuring that workflows are both scalable and reproducible. The success of repositories like nf-co.re \cite{ewels_nf-core_2020} illustrates the practical benefits of integrating reproducibility directly into workflow management systems.

\emph{REANA} is a newer workflow management system designed with a focus on reproducibility in data-intensive sciences, particularly high-energy particle physics \cite{chen_open_2019, simko_reana_2019}. Built on modern open-source tools like Kubernetes and Docker, REANA represents the state-of-the-art in workflow management. Mapping REANA to the generic workflow model is facilitated by its use of directed acyclic graph (DAG) models like CWL and Yadage, making the mapping process relatively straightforward. However, REANA’s simplicity, while robust, limits its flexibility in separating logical and physical workflow components, making reproducibility comparisons a manual task.
\subsubsection{Concrete mapping to \ac{DALiuGE}}\label{sec:daliugemapping}
\ac{DALiuGE} \citep{wu_daliuge:_2017}, \citep{wang_summit:2019} supports workflow management at \ac{SKA} scale.
The unique data-driven design, decentralized execution, and careful share-nothing architecture enable this.
Moreover, \ac{DALiuGE} strictly separates logical workflow design from physical workflow execution and thus allows astronomers to design workflows without intimate component knowledge and, more importantly, facilitates the broad reuse of logical workflows.
For this study, we have implemented the generalized ideas described in this work in \ac{DALiuGE} since we have complete code control and in-depth knowledge of the code structure. We, therefore, further elaborate on the workings of \ac{DALiuGE} and its mapping to the generic workflow model. \ac{DALiuGE} specifies workflows as graphs of components and separates workflows into two main layers, logical and physical, and then sub-divides each of those into template and non-template versions.
Workflows are developed as logical templates comprised of exclusively logical components and control structures.
Scientists will design a \ac{LGT} several months or years before the actual processing and for extensive reuse. \ac{LGT}s expose configuration options to make them more specific for concrete execution.
A \ac{LG} is a \ac{LGT} with a concrete configuration.
\ac{LG} specification occurs on the order of weeks before execution.
\ac{DALiuGE} then unrolls the \ac{LG} into a \ac{PGT}, resolving control structures and presenting the net computation required.
This process is unique for each \ac{LG}, but the result can vary depending on the translator algorithm used.
A physical graph template is a directed acyclic graph (DAG), a common representation for computational workflows \citep{ferreira_da_silva_characterization_2017}.
During the scheduling process, \ac{DALiuGE} partitions a \ac{PGT} into sub-graphs based on available compute resources, resulting in a \ac{PG}. This step is performed just before execution since it maps the physical workflow components to the available resources.
Each component of the DAG is a fully specified computational task or data artifact.
Finally, the \ac{DALiuGE} daemon triggers the first physical components in a graph, but after, components communicate with each other in a fully decentralized manner.
Components can store runtime-specific information such as log files or data summaries, which together form a \ac{RG} after graph execution has been completed. 

The first column of Table \ref{tab:workflowmappings} summarizes the mapping of the generic workflow model to DALiuGE components.
This mapping makes implementing reproducibility testing mechanisms feasible, and the already present distinction between ahead-of-time `logical' workflow design and runtime `physical' workflow execution makes mapping and subsequent implementation straightforward.
However, including a generic model and candidate mappings to other workflow management systems encourages further work.

\section{Reproducible Workflows}\label{sec:reproworkflows}
Contemporary science is complex; variability in the design and execution of complex data analysis pipelines can be scientifically significant \citep{botvinik-nezer_variability_2020}.
Most attempts to ensure workflows are reproducible involve analyzing artifact provenance, describing precisely how pieces of data came to exist \citep{celebi_towards_2020}.
We build on this lineage, but instead of designing provenance capture to describe the entire technical process, we formulate reproducibility tests, singling out parts of workflow executions expected to be kept constant between executions.
Doing so allows for flexible workflow characterization and does not limit this testing-based approach to a specific workflow management system, a critical conceptual requirement for the multi-decade-spanning SKA project and the broader scientific field.
Exploiting the structured nature of workflows provides constant time provenance comparison, which is suitably efficient for deployment at extreme data scales.

\subsection{Reproducibility Tenets}
Melding computing into satisfyingly reproducible science is a well-discussed and unsolved problem \citep{baker_1500_2016, ioannidis_meta-research_2018, gundersen_fundamental_2020}.
Current efforts range from asserting standards on individual pieces of code \citep{benureau_re-run_2018}, enacting well-defined data management protocols \citep{wilkinson_fair_2016, perez_systematic_2018}, containerizing whole workflows \citep{simko_reana_2019}, measures on behalf of journal publishers \citep{noauthor_ensuring_2020}, and conference organizers \citep{noauthor_transparency_2020} and careful provenance production and collection \citep{celebi_towards_2020}.
While certainly practical, we propose that forward-facing measures insufficiently integrate computing into science; they are ultimately empirical and require an inductive step to evaluate their efficacy.
The proposed approach demarcates implementation from abstract computational ideas, data and method reproduction, workflow similarities between workflow management systems, and any combination of these traits.
We present seven reproducibility tenets formulated as workflow component invariants, each defining a testable assertion rather than a series of prescribed rules.
For each tenet, we provide an unambiguous definition, an explanatory UML diagram based on the generic workflow model, a justification, and a few motivating example scenarios, providing practical motivation behind our definition.

\begin{table}[htbp]
\centering
\caption{Reproducibility tenets summarized as workflow component invariants.}
\resizebox{\columnwidth}{!}{%
\begin{tabular}{lcccc}
\hline
                                                                    & Log. Tasks         & Phys. Tasks         & Data Artifacts   & Scientific Information      \\ \hline
Rerun                                                               & \checkmark &            &           &         \\
Repeat                                                              & \checkmark & \checkmark &           &          \\
Recompute															&			 & \checkmark & 		  &         \\
Reproduce                                                           &            &            &  & \checkmark        \\
\begin{tabular}[c]{@{}l@{}}Scientific Replication\end{tabular}    & \checkmark &            &  & \checkmark         \\
\begin{tabular}[c]{@{}l@{}}Computational Replication\end{tabular} & 			 & \checkmark & \checkmark & \checkmark        \\
\begin{tabular}[c]{@{}l@{}}Total Replication\end{tabular} 		& \checkmark & \checkmark & \checkmark & \checkmark        \\ \hline
\end{tabular}%
}
\label{tab:Rtenets}
\end{table}

We use the popular terminology to Rerun, Repeat, Recompute, Reproduce, and Replicate computational workflows as these terms capture intuitive concepts around scientific reproducibility \citep{cohen-boulakia_scientific_2017, benureau_re-run_2018, barba_terminologies_2018, chen_open_2019}.
Table \ref{tab:Rtenets} summarizes the definitions, and the following subsections provide detailed explanations of each tenet in addition to the formal UML diagram for each of them.
The diagrams are structurally identical to Figure \ref{fig:WorkflowModel}, with the items required to be kept identical to satisfy that particular tenet highlighted in blue.
Later, Section \ref{sec:meth:tenetdata} describes the precise provenance information included in a concrete mapping to \ac{DALiuGE} to satisfy each tenet.

\subsubsection{Rerun}
\begin{figure}[htbp]
    \centering
    \includegraphics[width=\linewidth]{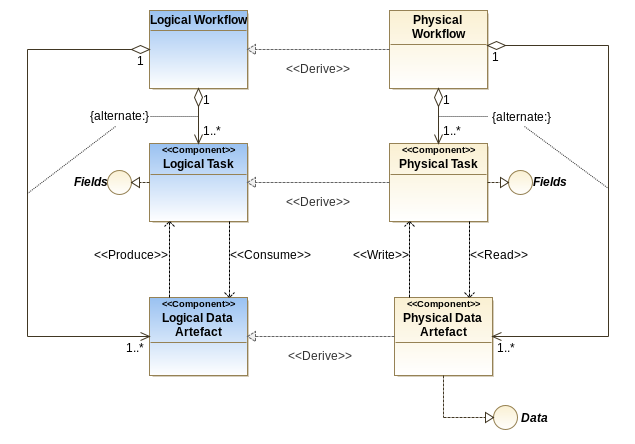}
    \caption{A UML workflow model with required identical components for workflow Reruns highlighted in blue.}
    \label{fig:wfm:rr}
\end{figure}

A workflow \emph{Reruns} another if they execute the same logical workflow; that is, their logical components and dependencies match.
Figure \ref{fig:wfm:rr} shows a version of the generic workflow model highlighting the elements required to be kept identical for workflow Reruns, in this case, logical components without instance-specific fields.
We intend that Reruns keep methods of computation identical in principle but open to multiple implementations.

Rerunning has been suggested to show that used tools are robust; Rerunning a computational workflow is analogous to using an identical algorithm independent of its implementation \citep{chen_open_2019}.
Asserting consistency only on the logical components in a workflow allows physical component changes.
This definition, in turn, allows changes in computing scale, parallelism patterns, hardware, data-encoding, or software implementation while still constituting a Rerun.
If all physical details are kept identical, a successful Rerun builds trust in the underlying tooling used to deploy a workflow, and a failed Rerun demonstrates system failure.
For some concrete examples, Rerunning workflows is useful when scaling up from a laptop to a cluster, for instance, or for workflows that, by design, deal with wildly varying data inputs, such as astronomy observatory ingest pipelines.

\subsubsection{Repeat}
\begin{figure}[htbp]
    \centering
    \includegraphics[width=\linewidth]{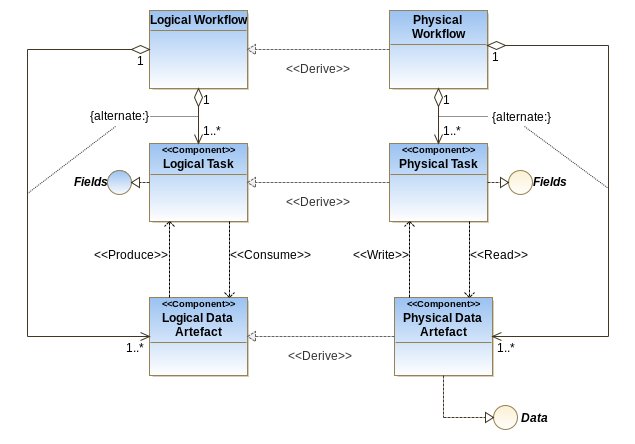}
    \caption{A UML workflow model with required identical components for workflow repetitions highlighted in blue.}
    \label{fig:wfm:rt}
\end{figure}

A workflow \emph{Repeats} another if they execute the same logical workflow and a principally identical physical workflow; their logical components, dependencies, and physical tasks match.
Figure \ref{fig:wfm:rt} shows which elements in the workflow model must be identical for two workflow executions to be considered repetitions.
The specificity of a `principally identical' physical workflow allows for a successful recovery from failed physical tasks (due to hardware or operating failure).
Repetitions are significantly more stringent than Reruns but leave the most specific details of physical tasks, such as the exact machine details these tasks run on and the content of the workflow's data artifacts.

Repetition is a relatively well-defined concept, and we generally follow intuition.
Repetition builds statistical power behind results \citep{chen_open_2019}.
By analogy to laboratory work, Repeated scientific computational workflows should match all components (workflow specification, data input, and runtime environment) \citep{cohen-boulakia_scientific_2017}.
This definition mostly follows existing interpretations but provides a concrete definition in which all logical and physical components match, excluding the runtime parameters of physical tasks.
Repeating a scientific computational workflow asserts the reuse of the logical and physical workflows, which, if successful, allows the bundling of data artifacts in concert.
For example, astronomical survey analysis runs on many patches of sky to create a full catalog, Repeating the same workflow but with different data inputs and outputs.
Repetition has also been known as methods reproducibility \citep{goodman_what_2016}.
Importantly, we leave the amount of information clarifying physical tasks as an environment and problem-dependent decision but insist that hardware details are not pertinent for repetition.

\subsubsection{Recompute}
\begin{figure}[htbp]
    \centering
    \includegraphics[width=\linewidth]{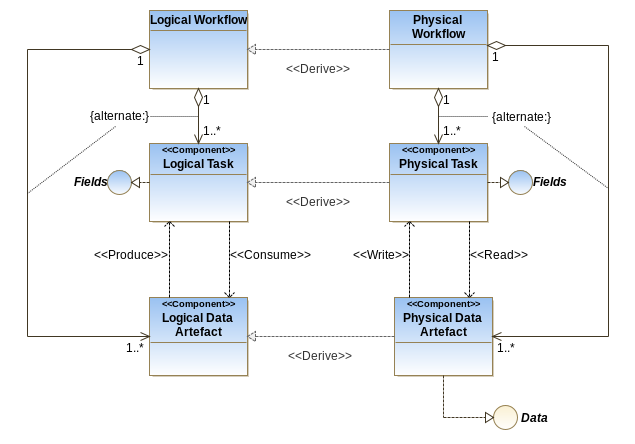}
    \caption{A UML workflow model with required identical components for workflow Recomputations highlighted in blue.}
    \label{fig:wfm:rc}
\end{figure}

A workflow \emph{Recomputes} another if they execute the same physical workflow; that is, their physical tasks and dependencies match precisely.
Figure \ref{fig:wfm:rc} shows the generic workflow model with all elements required for workflow Recomputation highlighted in blue.
The only element not required to be identical is the content of data artifacts.

Recomputation is a stricter interpretation of computational workflow repetition that requires identical physical tasks and hardware and is not fault-tolerant.
Workflow provenance, taken to an albeit extreme conclusion, motivates this definition; the effort required to Recompute a given workflow is immense and is likely impossible without total hardware control.
Some forms of repetition \citep{benureau_re-run_2018} are similar as they call principally for removing non-determinism.
In distributed computational workflows, the distribution itself can cause remarkably complex numerical non-determinism that is painstaking to correct \citep{iakymchuk_reproducibility_2020}, so Recomputation is a lofty endeavor.
However, Recomputation can be useful to demonstrate that a new computing facility is stable and isolate platform-dependent difficulties or errors, for example.

\subsubsection{Reproduce}
\begin{figure}[htbp]
    \centering
    \includegraphics[width=\linewidth]{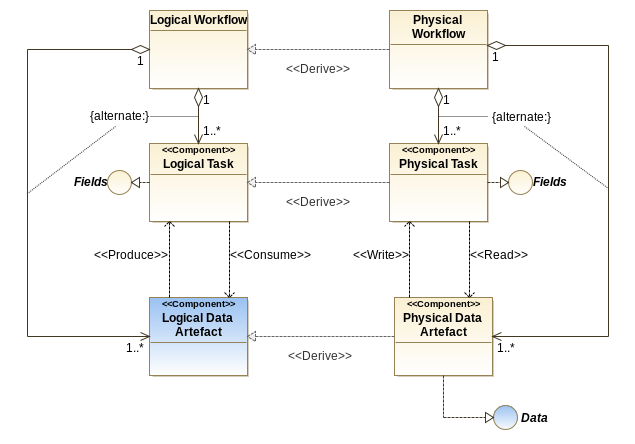}
    \caption{A UML workflow model with required identical components for workflow Reproductions highlighted in blue.}
    \label{fig:wfm:rp}
\end{figure}

A workflow \emph{Reproduces} another if their scientific information match.
Figure \ref{fig:wfm:rp} depicts the generic workflow model with components required for workflow reproduction highlighted in blue.
Only the logical description of data artifacts and the eventual data a workflow produces is required. The intention is to allow for entire methodologies to change, so long as the outputs of a workflow are maintained, it is a Reproduction, reaching the same end by any means. 

Scientific Reproduction is a widely discussed topic; our definition does not conflict with existing literature.
This definition of reproduction captures the net data flow a computational data-processing workflow consumes and produces.
Generally, reproducibility proper refers to reproducing experimental results \citep{fidler_reproducibility_2018}, an extremely difficult task in lab-based science \citep{lithgow_long_2017}.
Some define reproducible codes to be trivially executable and to produce published results \citep{benureau_re-run_2018}.
Our definition focuses solely on data in line with other interpretations \citep{chen_open_2019, barba_terminologies_2018} as opposed to more conceptual definitions based on conclusions \citep{cohen-boulakia_scientific_2017}.
The many interpretations of reproducibility present a challenge with many valid but incompatible definitions \citep{gundersen_fundamental_2020}.

Focusing on data artifacts provides an actionable approach to testing scientific reproducibility.
The formal reproduction of other teams' work and the reproduction of historical but impactful results are healthy for the wider academic community.
Moreover, reproduction alone permits absolute flexibility in methodology; demonstrating reproduction would be useful when establishing a radically different method to perform a known task, as is often the case when introducing machine learning methods, for example.
A tenet-testing-based approach is also compatible with reproducibility approaches involving constrained computing environments \citep{beaulieu-jones_reproducibility_2017} since we place no constraint on logical or physical task components.

\subsubsection{Replicate}
Replicability in science is a substantially debated concept with many interpretations \citep{fidler_reproducibility_2018, gundersen_fundamental_2020}.
We incorporate replication into our framework as combinations of previous tenets, creating three different but related standards that all constrain workflow tasks and data artifacts.

\begin{figure}[htbp]
    \centering
    \includegraphics[width=\linewidth]{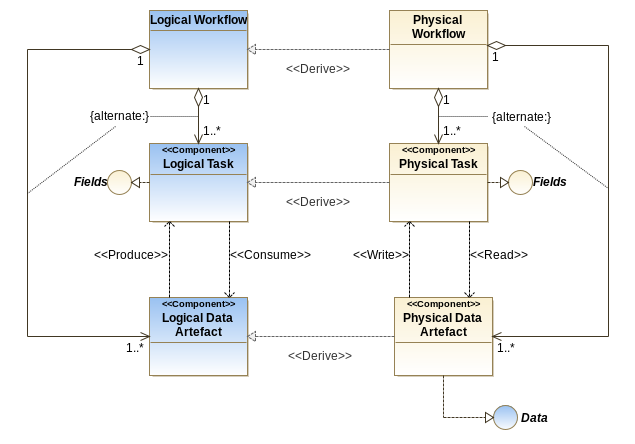}
    \caption{A UML workflow model with required identical components for workflow scientific replications highlighted in blue.}
    \label{fig:wfm:rpls}
\end{figure}

A workflow \emph{Scientifically Replicates} another if they \emph{Rerun} and \emph{Reproduce} each other.
Figure \ref{fig:wfm:rpls} shows the generic workflow model with components required for workflow scientific replication highlighted in blue. Logical components and the contents of data artifacts need to be identical. 
Scientific replication demonstrates that a particular method, represented as a logical workflow, produces a specific set of results.

This definition asserts that a principally identical computational workflow produces the same terminal data artifacts, in effect verifying that a logical workflow can produce a given set of results.
Our interpretations are reminiscent of what is termed conceptual interpretation \citep{fidler_reproducibility_2018} and of replicable code \citep{benureau_re-run_2018}.
Related but alternate views \citep{cohen-boulakia_scientific_2017} differ in that we do not focus on the conclusions a workflow implies.
For example, scientific replication is useful when updating library versions for a particular workflow, using different implementations for the same tasks, or moving to a new computing facility.

\begin{figure}[htbp]
    \centering
    \includegraphics[width=\linewidth]{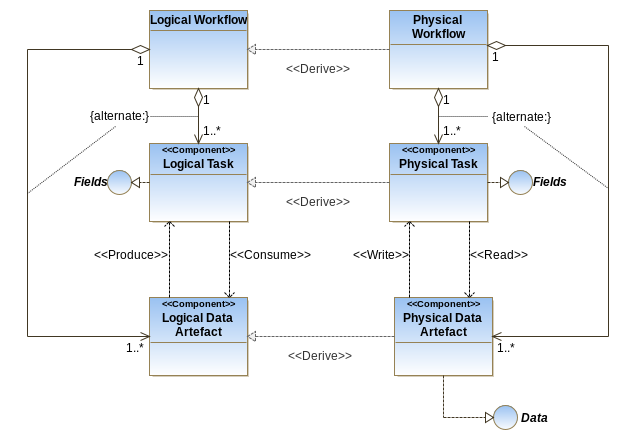}
    \caption{A UML workflow model with required identical components for workflow computational replications highlighted in blue.}
    \label{fig:wfm:rplc}
\end{figure}

A workflow \emph{Computationally Replicates} another if they \emph{Recompute} and \emph{Reproduce} each other and all data artifacts match.
Figure \ref{fig:wfm:rplc} contains the generic workflow model with components required for computational replication highlighted in blue.
It should be abundantly clear that computational replication is the most stringent possible form of replication permitted, as all elements must be identical.

This definition asserts that a precisely identical computational workflow produces the same data artifacts at every stage, which is the strictest interpretation of replication similar to the concept of direct replication \citep{fidler_reproducibility_2018}, and is closely related to workflow provenance discussed widely in computational workflow literature \citep{perez_systematic_2018}.
The information required to Replicate a workflow execution computationally is immense. It demonstrates that a specific workflow implementation on specific hardware yields a given set of results, building confidence in workflow implementation but not necessarily the underlying science; it would be challenging for an independent team to Totally Replicate a complex workflow.
Recomputation can confirm anomalous or unexpected results before further verification, for instance.

\begin{figure}[htbp]
    \centering
    \includegraphics[width=\linewidth]{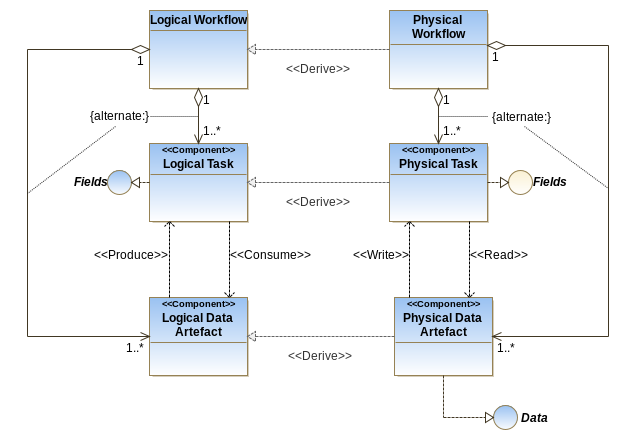}
    \caption{A UML workflow model with required identical components for workflow total replications highlighted in blue.}
    \label{fig:wfm:rplt}
\end{figure}

A workflow \emph{Totally Replicates} another if they \emph{Repeat} and \emph{Reproduce} each other and all data artifacts match.
Figure \ref{fig:wfm:rplt} contains the generic workflow model with components required for total replication highlighted in blue.

This definition asserts that a practically identical computational workflow produces the same data artifacts, including interim data, and are identical down to software, but not hardware, details.
Total replication is useful to certify a workflow as a `gold-standard' \citep{chen_open_2019} and is similar to the concept of results reproducibility \citep{goodman_what_2016}.
Despite computational replication being more exacting, we present this tenet last to reflect a balance between exacting replication, feasibility, and the scientific utility in total workflow replication.
For example, an integral processing task in radio astronomy requires separating unwanted radio frequency interference from astronomical observations, known as RFI flagging.
Not only is the correctness of this task important, but the speed of execution also matters significantly, as this task occurs before any science processing can commence.
Many possible methods exist spanning traditional algorithms and machine learning, and a performance claim around a new RFI flagging method could be strengthened by replicating the result independently on different hardware.
A total replication permits independent verification of results by a third party and balances scientific and computational replication.

\section{Reproducibility Signature Construction Methods}\label{sec:methods}
Having provided the theoretical background in the previous section, we outline a practical implementation in this section, allowing us to verify each reproducibility tenant. This includes the specific details behind the workflow signature mechanism and the information captured to build these signatures in DALiuGE, providing a single concrete mapping of the generic workflow model to a workflow management system and the reproducibility tenet definitions.
\subsection{Blockchain-Inspired Workflow Signature}
Building a dynamic workflow signature is central to our approach to reproducibility testing.
Moreover, at \ac{SKA} scale, these signatures need very efficient construction and comparison routines.
A significant challenge of managing computational workflows at this level of scale is the massive parallelism required to handle the volume of data these instruments produce.
As such, any workflow operation must handle extreme and dynamic scale in addition to online hardware failures, making robustness and networking efficiency central concerns.  
To this end, we take inspiration from blockchain technologies, borrowing the Merkle tree cryptographic data structure \citep{merkle_digital_1988}.
Information about each workflow model component, dictated by tenet definitions, is stored in a Merkle tree, the root of which provides a signature for that component.
Inserting these signatures into a blockchain-like hash-graph mirroring their connectivity in the original workflow creates a \ac{BlockDAG}.
The root of a Merkle tree constructed from the leaves of this \ac{BlockDAG} is a signature for the entire workflow.
Similar to how blockchains are cryptographically secure, subtle changes to component provenance information result in drastic changes to the final signature, but structuring this signature makes it efficient to identify where these changes occur.
Given two workflow BlockDAGs, one can walk through a topological sort of the nodes to find the first differing signature.
The information stored in each block by each workflow component changes for each reproducibility tenet. We collect the hashes of leaf nodes, insert those into a Merkle tree, and take the root as a signature for the entire workflow for a particular tenet.

We use Kahn's algorithm \citep{kahn_topological_1962} to traverse a workflow description in a topological order taking $\mathcal{O}(V+E)$ time where $V$ is the number of components (vertices) in the workflow and $E$ is the number of edges between all components.
Construction of each component Merkle tree takes $\mathcal{O}(|V|\,log\,|V|)$ time where $|V|$ is the volume of provenance information the component stores.
Inserting a component into the \ac{BlockDAG} takes $\mathcal{O}(d\,log\,d)$ time, where 
$d$ is the degree of that component.
Constructing the entire \ac{BlockDAG} takes $\mathcal{O}(V(D\,log,D)+E)$ time, where $D$ is the average vertex degree.
Construction of the final signature takes $\mathcal{O}(l\,log\,l)$ time, where $l$ is the number of leaves in the workflow graph.
When amortized against the time taken to execute a workflow in the first place ($\mathcal{O}(V + E)$), signature construction takes $\mathcal{O}(log\,l)$ time.
\subsection{Example Signature Construction}\label{sec:meth:tenetdata}
\begin{figure*}[!htbp]
    \centering
    \includegraphics[height=45mm]{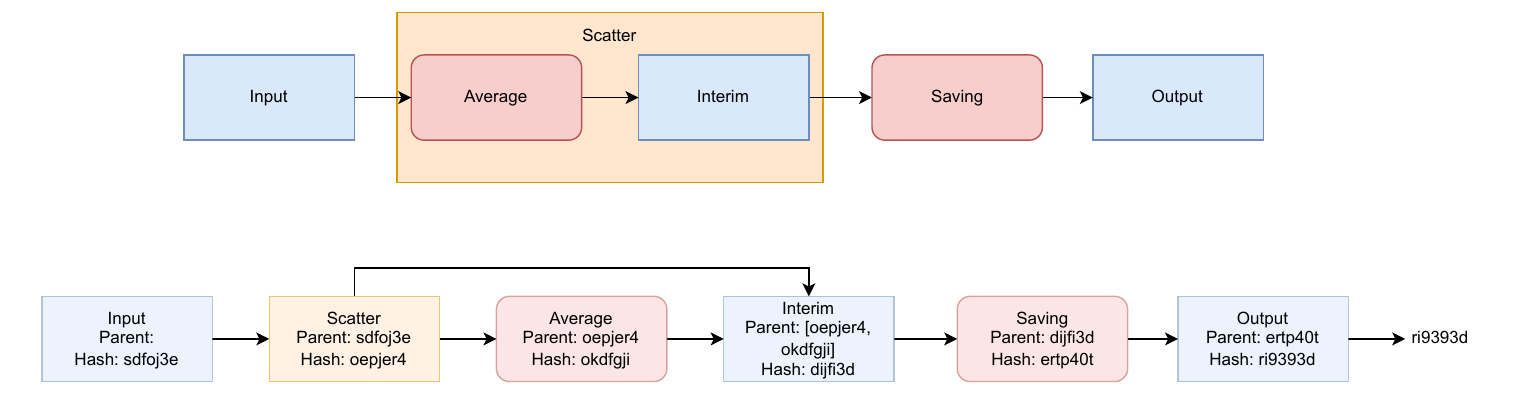}
    \caption{Example Logical Graph Template (above) and associated hash-graph (below). All components have a one-to-one relation in the BlockDAG. Red boxes represent tasks, blue boxes represent data artifacts, and yellow boxes represent control structures.}
    \label{fig:ExLGT}
\end{figure*}

\begin{figure*}[!htbp]
    \centering
    \includegraphics[height=45mm]{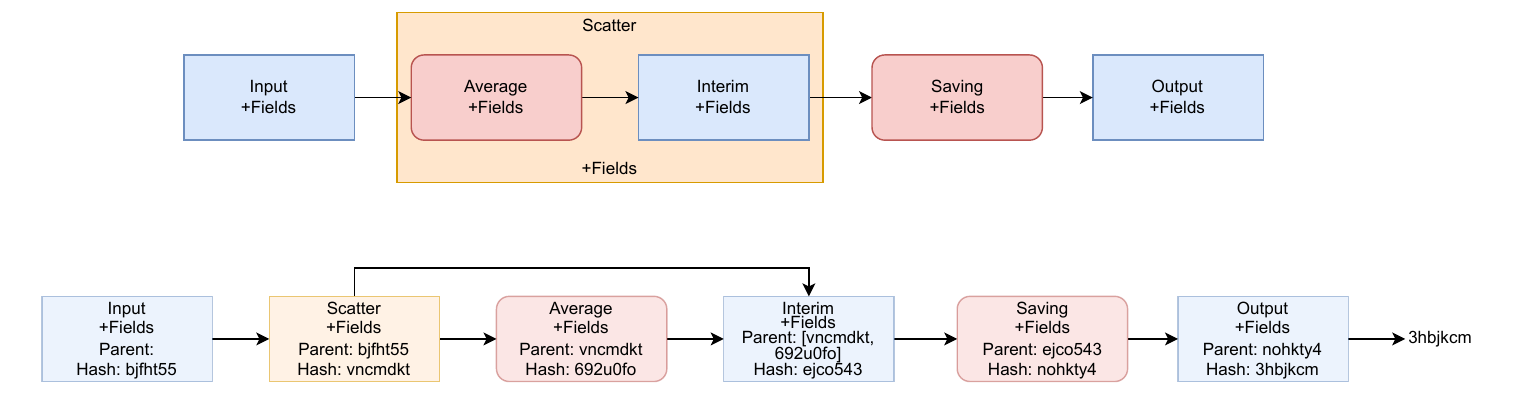}
    \caption{Example Logical Graph (above) and associated hash-graph (below). Here, the `fields' entry refers to arbitrary additional information required for a particular tenet. Boxes hold the same meaning as in Figure \ref{fig:ExLGT}.}
    \label{fig:ExLG}
\end{figure*}

\begin{figure*}[!htbp]
    \centering
    \includegraphics[height=55mm]{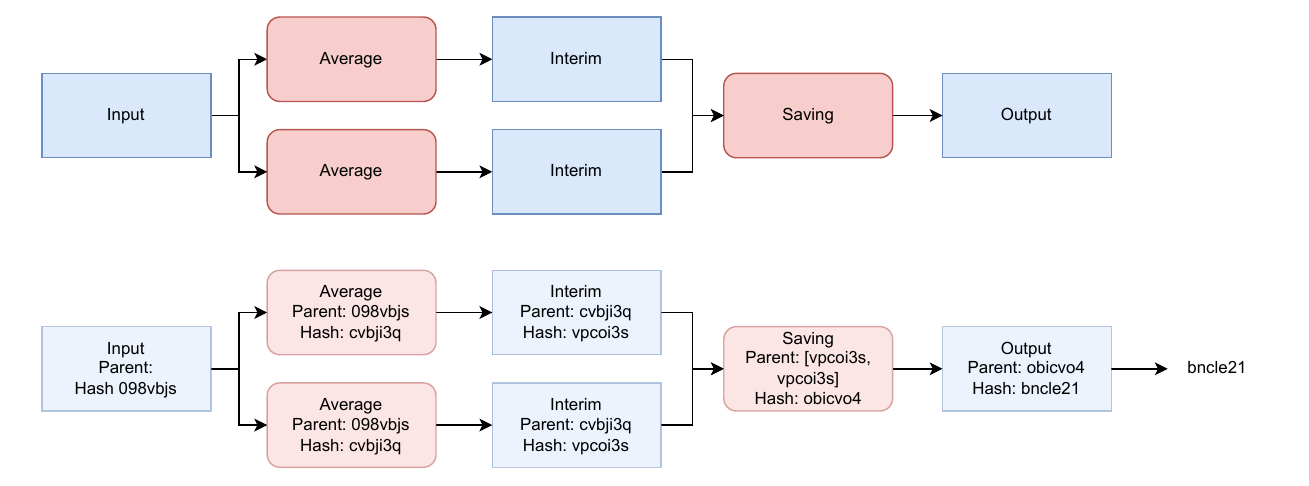}
    \caption{Example Physical Graph Template (above) and associated hash-graph (below). Control structures have been removed as the workflow has been unrolled. Boxes hold the same meaning as in Figure \ref{fig:ExLGT}.}
    \label{fig:ExPGT}
\end{figure*}

\begin{figure*}[!htbp]
    \centering
    \includegraphics[height=55mm]{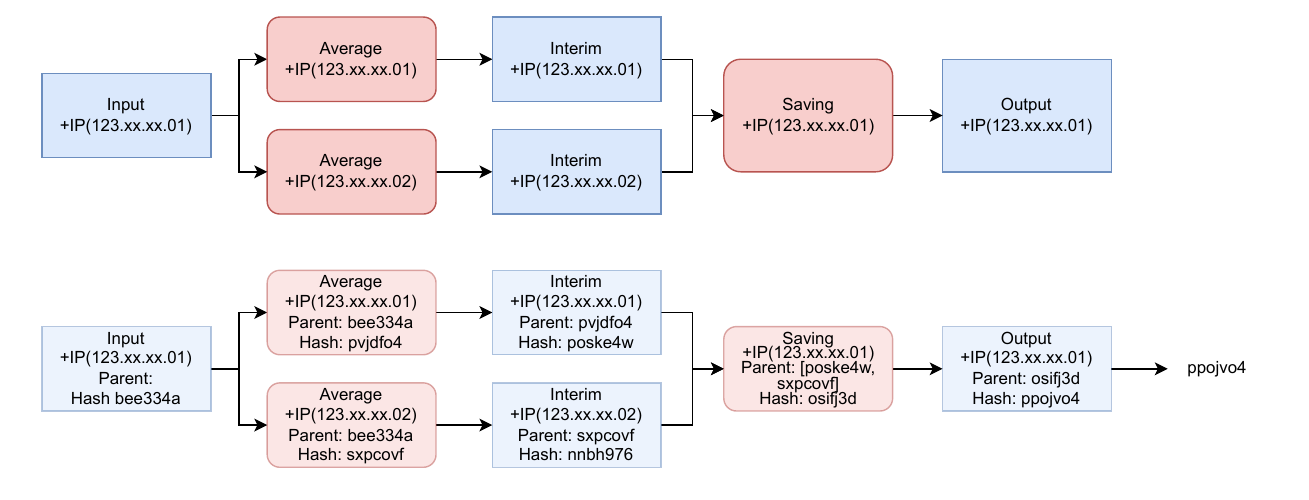}
    \caption{Example Physical Graph (above) and associated hash-graph (below). We include the IP address of a particular task's machine as an example of physical task specific information. Boxes hold the same meaning as in Figure \ref{fig:ExLGT}.}
    \label{fig:ExPG}
\end{figure*}

\begin{figure*}[!htbp]
    \centering
    \includegraphics[height=55mm]{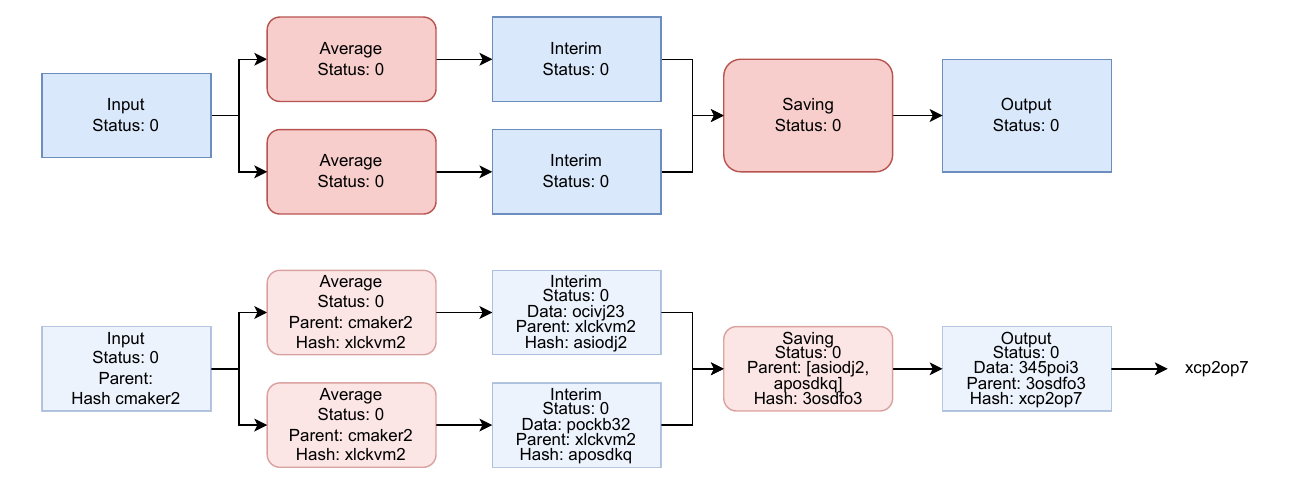}
    \caption{Example Runtime Graph (above) and associated hash-graph (below). The status field refers to the execution exit code as an example of runtime-specific information. Boxes hold the same meaning as in Figure \ref{fig:ExLGT}.}
    \label{fig:ExRG}
\end{figure*}

\begin{figure*}[!htbp]
    \centering
    \includegraphics[height=124mm]{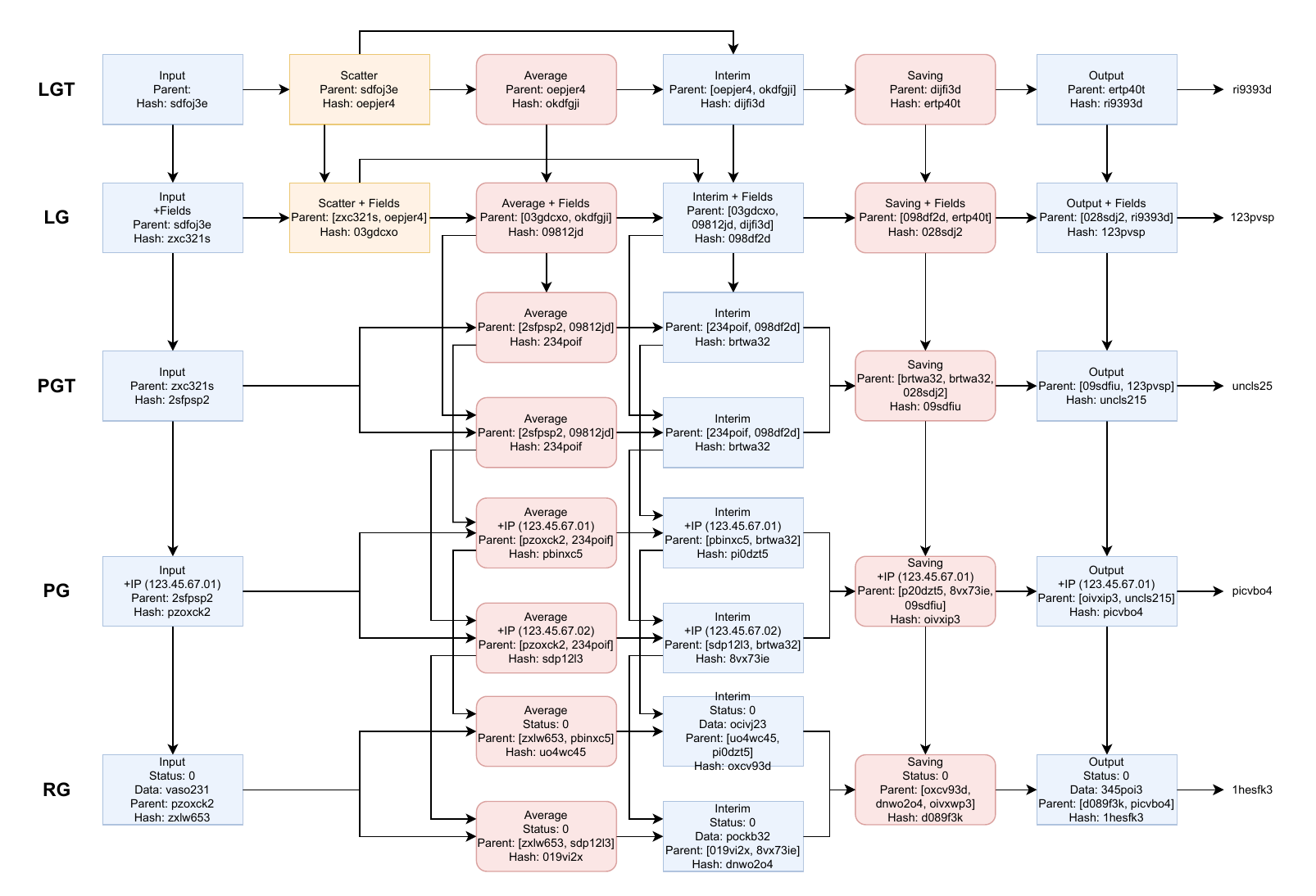}
    \caption{Example combined hash-graph. The relationships between components throughout their lifecycle, from logical to physical components, are explicit. While the final signatures of each DALiuGE workflow abstraction from LGT to RG are shown separately, in practice, we would hash these signatures together into a single final signature. Boxes hold the same meaning as in Figure \ref{fig:ExLGT}.}
    \label{fig:ExCOMB}
\end{figure*}

We elaborate on workflow signature construction through an existing simple example \cite{pritchard_using_2021}.
This simple example averages a series of numbers split across multiple tasks, combining the total average in a single process.
We present this example using terminology from the generic workflow model but in the phases designated in the life of a DALiuGE workflow, outlined in Section \ref{sec:daliugemapping}.
Figures \ref{fig:ExLGT} to \ref{fig:ExCOMB} provide hash-graphs of a workflow \ac{BlockDAG} over successive workflow phases under computational replication standards. Information about logical tasks, physical tasks, and logical and physical data artifacts are included to illustrate how this information comes together.
Figure \ref{fig:ExLGT} presents a logical graph template for a simple averaging workflow and its associated BlockDAG.
The original graph structure is maintained, each component generates a hash-value combining its parents' hash and its own specification.

A logical graph is a logical template with specified options.
Figure \ref{fig:ExLG} presents an example logical graph.
The structure has not changed, but more information is incorporated into the signatures, indicated by the `fields' entry, changing the resulting hash values.

Unrolling a logical template into a physical template resolves control structures, yielding a complete description of the computing and storage tasks required.
Physical graph templates are directed acyclic graphs. 
Figure \ref{fig:ExPGT} shows a physical graph template where DALiuGE scatters information across two branches.

Figure \ref{fig:ExPG} shows a physical graph representing a fully specified physical workflow.
In this example, physical tasks are spread across multiple machines in a cluster, resulting in principally identical tasks holding different field information (IP address, for example).

After runtime, the hash graph contains runtime information such as task completion status and data-artifact hashes.
Figure \ref{fig:ExRG} shows this runtime graph.

Finally, Figure \ref{fig:ExCOMB} combines the hash-graphs of all workflow components into a final signature.
Visualizing more complex workflows does become arduous. However, we aim to show precisely how a signature is created from Logical and physical workflow components.

\FloatBarrier

\subsection{Tenet Data}
We now elaborate on what specific information each component type in DALiuGE captures for each tenet.
The fields we select are ultimately an interpretation of the tenet definitions and, as such, are open to adjustment, especially when considering alternate workflow management systems or native software.
The fields selected represent the mapping of the generic workflow model to DALiuGE components, which we have aggregated into `Application' drops, `Data' drops, and `Control' drops for brevity.

Some additional information about DALiuGE is required to understand all fields present.
`InPorts' and `OutPorts' represent the DALiuGE-specific abstraction for component dataflow connections.
`Node' refers to the logical (integer-valued) or physical (IP address) machine a component will execute on.
Status refers to execution status, trace refers to a stack trace of that component's execution and not all fields are present in all components, as is the case for Filenames in application components, for example.
`Type' is an increasingly detailed field as a workflow moves closer to execution. In an LGT, `type' will refer to a generic tool like bash or Python; in a \ac{PGT}, `type' will refer to a specific toolchain like a Python version.
`Name' refers to the exact application name, such as a specific Python script. 
\paragraph{Rerun}
Rerunning two workflow executions requires matching their logical workflow components.
A DALiuGE mapping implies that logical component details must match.
Unsurprisingly, most information captured in this tenet is at the logical layers, only tracking the types and ultimate execution status of each physical component.
Table \ref{tab:RR} contains a summary of the fields each component type stores in its signature when included in Rerun BlockDAGs.
\begin{table}[!htbp]
\centering
\caption{Information stored by DALiuGE components to assert workflow Reruns for different components.}
\resizebox{\columnwidth}{!}{%
\begin{tabular}{ccccccc}
Component-type      & LGT                                                               & LG                                                                 & PGT & PG & RG     \\ \hline
Application    & \begin{tabular}[c]{@{}c@{}}Type\\ InPorts\\ OutPorts\end{tabular} & -  & \begin{tabular}[c]{@{}c@{}}Type\\ Name\end{tabular}   & -  & Status \\ 
Data  & ``                                                                 & ``  & \begin{tabular}[c]{@{}c@{}}Type\\ Storage-Name\end{tabular} & ``  & Status \\ 
Control (misc) & ``                                                                 & ``  & \begin{tabular}[c]{@{}c@{}}Type\end{tabular} & ``  & -     
\end{tabular}
\label{tab:RR}%
}
\end{table}
\paragraph{Repeat}
In addition to the information stored for asserting Reruns, Repetition stores additional information about component fields (adjustable parameters).
Table \ref{tab:RT} summarises the fields present in each component type when included in Repeat BlockDAGs.

\begin{table}[!htbp]
\centering
\caption{Information stored to assert workflow repetitions for different components.}
\resizebox{\columnwidth}{!}{%
\begin{tabular}{ccccccc}
Component-type      & LGT                                                               & LG                                                                                                                                             & PGT & PG & RG     \\ \hline
Application    & \begin{tabular}[c]{@{}c@{}}Type\\ InPorts\\ OutPorts\end{tabular} & \begin{tabular}[c]{@{}c@{}}Num-CPUs\\ Fields\end{tabular} & \begin{tabular}[c]{@{}c@{}}Type\\ Name\end{tabular}         & -  & Status \\
Data  & ``                                                                & \begin{tabular}[c]{@{}c@{}}Data-Volume\\ Fields\end{tabular}   & \begin{tabular}[c]{@{}c@{}}Type\\ Storage-Name\end{tabular}  & `` & Status \\ 
Control (misc) & ``                                                                & Fields                                                                     & \begin{tabular}[c]{@{}c@{}}Type\end{tabular}                  & `` & -     
\end{tabular}
\label{tab:RT}%
}
\end{table}

\paragraph{Recompute}
Recomputation captures significantly more details of component field values and execution environment. 
This definition and workflow mapping implies that application drops must match exactly.
Table \ref{tab:RC} summarises the information we capture in testing workflow Recomputation.
In addition to the component details, we store the logical partitioning of components to machine resources and an execution trace for each processing component.

\begin{table}[!htbp]
\centering
\caption{Information stored to assert workflow Recomputations for different components.}
\resizebox{\columnwidth}{!}{%
\begin{tabular}{ccccccc}
Component-type      & LGT                                                               & LG                                                                                                                                              & PGT                                                   & PG                                                          & RG                                                             \\ \hline
Application    & \begin{tabular}[c]{@{}c@{}}Type\\ InPorts\\ OutPorts\end{tabular} & \begin{tabular}[c]{@{}c@{}} Num-CPUs\\ Fields\end{tabular} & \begin{tabular}[c]{@{}c@{}}Type\\ Rank\\ Name\\Node\\ Island\end{tabular} & \begin{tabular}[c]{@{}c@{}}Node-IP\\ Island-IP\end{tabular} & \begin{tabular}[c]{@{}c@{}}Status\\ Trace\end{tabular} \\ 
Data  & ``                                                                & \begin{tabular}[c]{@{}c@{}}Data-Volume\\ Filenames\\ Fields\end{tabular}   & \begin{tabular}[c]{@{}c@{}}Type\\ Rank\\ Storage-Name\end{tabular}                                                    & ``                                                          & Status                                                         \\ 
Control (misc) & ``                                                                & Fields                                                                     & \begin{tabular}[c]{@{}c@{}}Type\\ Rank\end{tabular}                                                                  & ``                                                          & -                                                             
\end{tabular}
\label{tab:RC}%
}
\end{table}

\paragraph{Reproduce}
Reproduction significantly differs, primarily focusing on data components.
We store information to assert that data components are configured identically and store the same data summary.
The mapping to DALiuGE requires logical data components and data drops to match their specification and content between workflow executions.
How components generate their data summary is a subtlety worth noting; in our case, this is a hash of the data contents.
In addition to a simple content hash, this content signature could be implemented by calculating statistical properties or other domain-specific characteristics, such as existing hashes in file headers.
Table \ref{tab:RP} summarises the fields different components store when included in a Reproduction BlockDAG.

\begin{table}[!htbp]
\centering
\caption{Information stored to assert workflow Reproductions for different components.}
\resizebox{\columnwidth}{!}{%
\begin{tabular}{ccccccc}
Component-type      & LGT  & LG                                                                 & PGT & PG & RG           \\ \hline
Application    & Type & -  & -   & -  & -            \\ 
Data  & ``   & `` & \begin{tabular}[c]{@{}c@{}}Type\\ Storage-Name\end{tabular}  & `` & \begin{tabular}[c]{@{}c@{}}Status\\Data-Summary\end{tabular}\\ 
Control (misc) & ``   & `` & -   & `` & -           
\end{tabular}
\label{tab:RP}%
}
\end{table}

\paragraph{Replicate Scientific/Computational/Total}
Replication information is straightforwardly a combination of the Rerun, Recompute, or Repetition information (respectively) and Reproduction fields.
Tables \ref{tab:RPLS}, \ref{tab:RPLC} and \ref{tab:RPLT} summarise the fields required for scientific, computational, and total replication BlockDAGs, respectively.

\begin{table}[!htbp]
\centering
\caption{Information stored to assert workflow Scientific-Replication for different components.}
\resizebox{\columnwidth}{!}{%
\begin{tabular}{ccccccc}
Component-type      & LGT                                                               & LG                                                                 & PGT & PG & RG           \\ \hline
Application    & \begin{tabular}[c]{@{}c@{}}Type\\ InPorts\\ OutPorts\end{tabular} & -  & \begin{tabular}[c]{@{}c@{}}Type\\ Name\end{tabular}           & -  & Status       \\ 
Data  & ``                                                                & `` & \begin{tabular}[c]{@{}c@{}}Type\\ Storage-Name\end{tabular}   & `` & \begin{tabular}[c]{@{}c@{}}Status\\Data-Summary\end{tabular} \\ 
Control (misc) & ``                                                                & `` & \begin{tabular}[c]{@{}c@{}}Type\end{tabular}                  & `` & -           
\end{tabular}
\label{tab:RPLS}%
}
\end{table}

\begin{table}[!htbp]
\centering
\caption{Information stored to assert workflow Computational-Replication for different components.}
\resizebox{\columnwidth}{!}{%
\begin{tabular}{ccccccc}
Component-type      & LGT                                                               & LG                                                                                                                                              & PGT                                                   & PG                                                          & RG                                                            \\ \hline
Application    & \begin{tabular}[c]{@{}c@{}}Type\\ InPorts\\ OutPorts\end{tabular} & \begin{tabular}[c]{@{}c@{}}Num-CPUs\\ Fields\end{tabular}          & \begin{tabular}[c]{@{}c@{}}Type\\ Rank\\ Name\\Node\\ Island\end{tabular} & \begin{tabular}[c]{@{}c@{}}Node-IP\\ Island-IP\end{tabular} & \begin{tabular}[c]{@{}c@{}}Status\\ Trace\end{tabular}        \\ 
Data  & ``                                                                & \begin{tabular}[c]{@{}c@{}}Data-Volume\\ Filenames\\ Fields\end{tabular}   & \begin{tabular}[c]{@{}c@{}}Type\\ Rank\\ Storage-Name\end{tabular}                                                    & ``                                                          & \begin{tabular}[c]{@{}c@{}}Status\\ Data-Summary\end{tabular} \\ 
Control (misc) & ``                                                                & Fields                                                                     & \begin{tabular}[c]{@{}c@{}}Type\\ Rank\end{tabular}                                                                    & ``                                                          & -                                                            
\end{tabular}
\label{tab:RPLC}%
}
\end{table}

\begin{table}[!htbp]
\centering
\caption{Information stored to assert workflow Total-Replication for different components.}
\resizebox{\columnwidth}{!}{%
\begin{tabular}{ccccccc}
Component-type      & LGT                                                               & LG                                                                                                                                             & PGT & PG & RG                                                            \\ \hline
Application    & \begin{tabular}[c]{@{}c@{}}Type\\ InPorts\\ OutPorts\end{tabular} & \begin{tabular}[c]{@{}c@{}}Num-CPUs\\ Fields\end{tabular} & \begin{tabular}[c]{@{}c@{}}Type\\ Rank\\ Name\end{tabular}           & -  & Status                                                        \\ 
Data  & ``                                                                & \begin{tabular}[c]{@{}c@{}}Data-Volume\\ Fields\end{tabular}   & \begin{tabular}[c]{@{}c@{}}Type\\ Storage-Name\end{tabular}  & `` & \begin{tabular}[c]{@{}c@{}}Status\\ Data-Summary\end{tabular} \\ 
Control (misc) & ``                                                                & Fields                                                                     & \begin{tabular}[c]{@{}c@{}}Type\\ Rank\end{tabular}                  & `` & -                                                            
\end{tabular}
\label{tab:RPLT}%
}
\end{table}

Establishing the method used to create a workflow signature and the exact mapping of fields from the generic workflow model to DALiuGE components under different tenet definitions makes the process of workflow signature generation clear.
\section{Demonstration}\label{sec:demonstration}
In this section, we fully demonstrate the reproducibility tenets and BlockDAG signature-based testing mechanism.
First, we describe a lowpass filter workflow implemented in DALiuGE and a second workflow that tests the effectiveness of a particular lowpass filter.
Then, we execute each workflow ten times with different implementations of the lowpass filter element, constructing signatures for each of the reproducibility tenets as described previously.
For each tenet, we comment on what the resulting signatures reveal, finding that subtle numerical differences that would otherwise be elusive are obvious when comparing signatures.

\subsection{Demonstrative Lowpass Workflow}\label{sec:fftwflow}
\begin{figure*}[htbp]
    \centering
    \includegraphics[width=\textwidth]{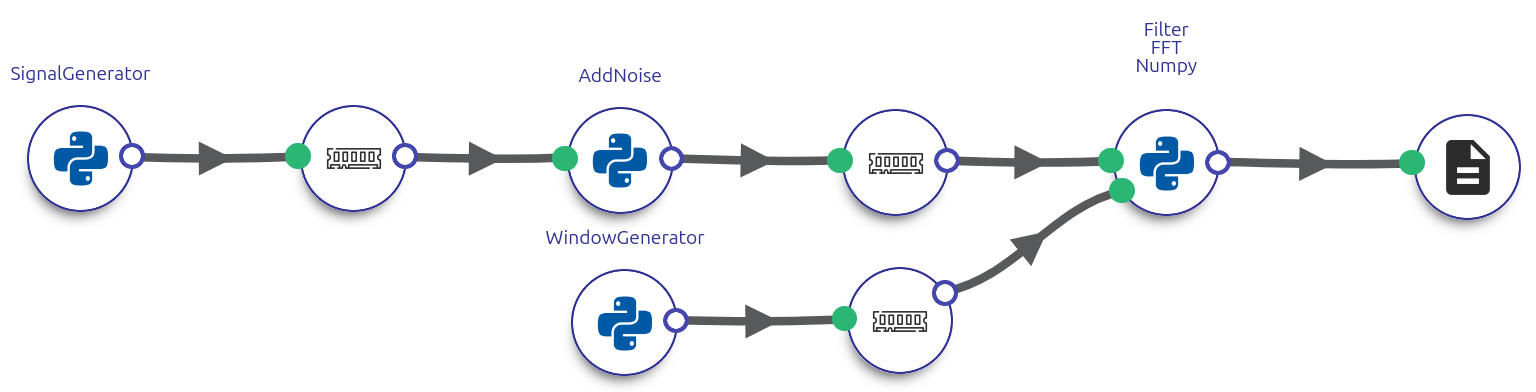}
    \caption{Generic lowpass filter workflow. The filter component is swapped out for specific implementations, but the rest of the workflow remains the same. Circles with Python logos represent Python tasks, and the RAM logos represent data storage in memory. The file icon represents a result written to storage and is accessible after workflow execution.}
    \label{fig:fftworkflow}
\end{figure*}

We first present two workflows implemented in DALiuGE and designed in Eagle \citep{EAGLE}, a graphical workflow designer for DALiuGE.
The first workflow generates a pure sine wave, injects random Gaussian noise, and then attempts to filter the result with a lowpass filter (based on convolution with a Hann window).
While structurally simple, each component is highly configurable and we implement the core filter component with four different methods; a NumPy \citep{harris_array_2020} pointwise brute-force convolution filter for reference, a NumPy \ac{FFT} filter, a FFTW \citep{henry_gomersall_pyfftw_2020} \ac{FFT} filter and PyCuda \citep{klockner_pycuda_2012} GPU based \ac{FFT} filter.
Figure \ref{fig:fftworkflow} contains a single rendering of this workflow.
This workflow highlights numerical differences between \ac{FFT} implementations, a fundamental operation in astronomical sciences and wider scientific computing.

\begin{figure*}[htbp]
    \centering
    \includegraphics[width=\textwidth]{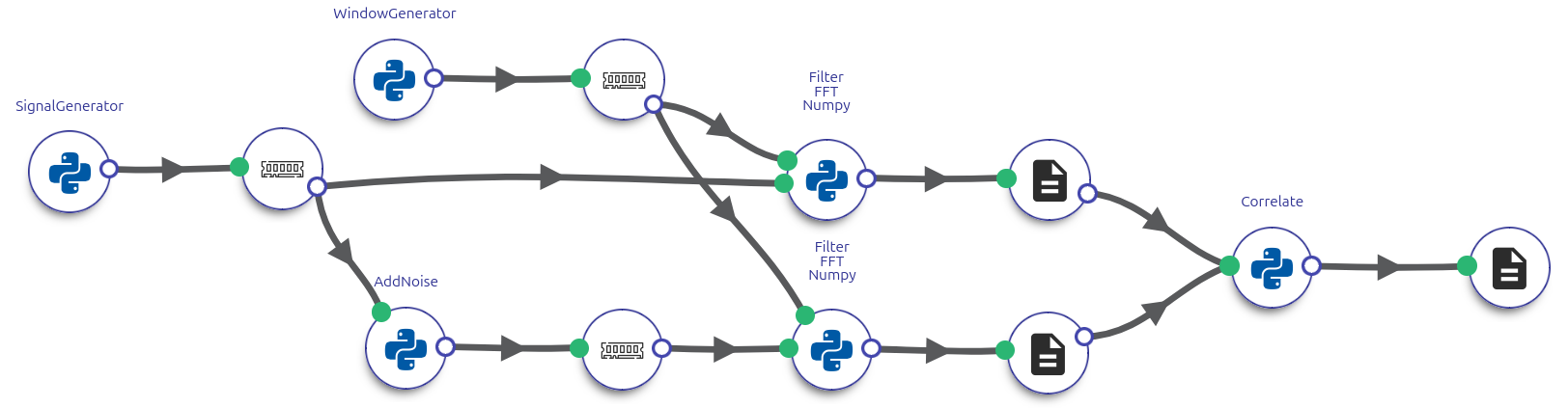}
    \caption{Workflows including the filter task and \ac{NCC} calculation. Symbols are as in Figure \ref{fig:fftworkflow}.}
    \label{fig:fftworkflowncc}
\end{figure*}

In addition to the basic workflow, we construct a second workflow that computes the Normalized Cross-Correlation \ac{NCC} \citep{ingle_statisical_2005} between the original noisy signal and the recovered filtered signal divided by the length of the signal.
The resulting metric effectively gives the probability that any filtered sample matches its corresponding sample in the original signal.
Figure \ref{fig:fftworkflowncc} contains a single rendering of this workflow using EAGLE, the \ac{DALiuGE} graphical workflow editor \cite{EAGLE}.
This metric is relative, and therefore, we use the length of the signal to determine the numerical precision required to compare the different methods as:
\begin{equation}
    2^{\lceil log_2(L) \rceil}.
\end{equation}
The signal length is $512$ in all trials, so we require three digits of precision in the NCC metric.
We execute each basic and extended workflow ten times, changing the random seed for the noise generator to the trial number in both cases.

To demonstrate this methodology as a practical means of comparing workflow designs and implementations, we provide example workflow executions designed to highlight the reproducibility tenets and real-world examples of where they would apply to astronomical interferometric data processing.
All tenet signatures for all referenced workflow trials are available online \citep{pritchard_lowpass_trials_2022}.
Running multiple trials of each workflow, changing only the random noise introduced to the base signal, and comparing tenet hashes provide a straightforward way to explore what they uncover.
Table \ref{tab:res:plain} in \ref{app:results} contains the truncated hashes of all plain workflow executions, omitted here for brevity.
Table \ref{tab:res:ncc} contains the truncated hashes of all NCC test workflow executions, omitted here for brevity.

We now outline example use cases for each tenet:
\paragraph{Rerun} We compare the Rerun hashes of each standard workflow execution and of each of the NCC workflows.
Successful Reruns intend to establish trust in the workflow management system itself, as mentioned previously.
While each workflow uses different components, they are all implemented in Python and should all be Reruns, which we find quite straightforwardly to be the case, shown by the identical Rerun hashes.
\paragraph{Repeat} We compare the Repeat hashes of each extended workflow trial and find that the Repeat hashes of each filter method match in their signature.
Trials that match can effectively have their results bundled together to make a subsequent claim.
We can confidently conclude that each NCC workflow execution for each filter method did execute the same workflow components, independent of the data they ingested or output.
\paragraph{Recompute} We can compare the Recompute hashes of each workflow trial and find that all trials have a different Recomputation hash.
Without an identical computing environment, one would not expect any workflow execution here to be a Recomputation.
As defined, demonstrating workflow Recomputation is an extraordinarily difficult task.
\paragraph{Reproduce} For each Repeated trial, that is, bundling signatures based on filter implementation, we see that the individual filter workflows all produce different reproduction signatures.
This is to be expected since each method receives a different signal as input owing to the random noise injected, producing subtly different outputs.
However, when comparing the Repeated trials for the NCC workflows, we see that some, but not all, workflow executions are Reproductions.
Since the final output of the NCC workflows is effectively a reconstruction error and despite the input being different for each trial, all executions fall into one of two signatures: `8fd08' or `bf29b'.
Upon further inspection, we find a slight but potentially impactful difference of 0.001.
This test exhibits how building results post-processing into workflows directly allows for the precise testing of highly abstract ideas (e.g., the performance of a filter implementation).
\paragraph{Replicate Scientific/Computational/Total} For each type of replication, we compare the hashes of the extended workflow trials.
One would expect some Repeat trials of the same method to be scientific replicas since the performance of a particular filter should be consistent.
A computational replica would be highly surprising, and establishing total replication verifies that a particular method produces a particular result.
We find that workflows that are Reproductions are also scientific and total Replications.
By formally demonstrating that different filtering executions produce different results, regardless of what filtering method is used, but also showing that the final error in filtering is consistent across each method tested, we can assert with confidence an equivalence between the FFTW, Cuda FFT, Numpy FFT, and Numpy Pointwise workflows.

\subsection{Compute Environment}
All trials were executed on a machine with an Intel Core 17-10875H, Nvidia RTX 2080 Super (Mobile), and 64GB of RAM.
Experiments were run with DALiuGE version 2.3.2, EAGLE version 4.5.2, and lowpass components version 0.2.

\section{Conclusion}\label{sec:conclusion}
Computational workflows power contemporary data-intensive sciences, and play an increasingly important role in astronomical sciences in particular.
Research engineers adopt industry-standard software development methodologies to build scientific trust in these complex software systems to implement and test the software.
We provide an additional methodology for testing and verifying consistency between and across execution runs using either the same or varying implementations of specific algorithms.
We formally define seven reproducibility tenets based on a generic workflow UML model.
We then consider specific provenance information required to satisfy each definition and construct a hash-graph-based signature mechanism termed BlockDAGs.
We provide a concrete example implementation of this mechanism in \ac{DALiuGE} connecting the UML scale-agnostic workflow model, reproducibility tenet definitions, which inform provenance information collection, and workflow signature mechanism together.
A demonstrative lowpass filter example shows that this method asserts the scientific equivalence of four implementations of the same basic workflow while simultaneously identifying their output-data differences.
Workflow management systems guide workflow execution and design, and this method of testing the scientific equality between workflow executions allows us to verify that these systems deliver their promised benefits.
Additional considerations, such as adding formal provenance data to the component hashes and assessing the impact of hardware failure on large scientific workflow instances, would further verify the efficacy of this method.
Implementing a BlockDAG mechanism in other workflow management systems would allow for workflow verification between workflow management systems.
By providing a test-driven method to determine the scientific equivalence between workflow executions, we provide verifiable evidence that next-generation science instruments, such as the \ac{SKA}, produce high-quality, reproducible results.

\section{Software and data availability}
All components used in this study are publicly available as a Python package \citep{pritchard_lowpass_components_2022}. The workflows themselves \citep{pritchard_lowpass_graphs_2022}, and all data products from each workflow trial execution are available online \citep{pritchard_lowpass_trials_2022}. The \ac{DALiuGE} system \citep{daliuge_soft} and the {DALiuGE} graph editor {EAGLE} \citep{EAGLE} are open source and available as well.

\section*{CRediT authorship contribution statement}
\textbf{Nicholas J. Pritchard}: Conceptualization, Methodology, Software, Validation, Investigation, Data Curation, Writing - Original Draft, Writing - Review \& Editing.
\textbf{Andreas Wicenec}: Conceptualization, Resources, Writing - Review \& Editing, Supervision.
\section*{Declaration of competing interest}
The authors declare that they have no known competing financial interests or personal relationships that could have appeared to influence the work reported in this paper.
\section*{Data availability}
Data will be made available on request.
\section*{Acknowledgements}
This research was supported by an Australian Government Research Training Program (RTP) Scholarship.

\appendix
\section{Workflow trial signatures}
\label{app:results}
Here, we present truncated signatures output from all trial runs referred to in Section \ref{sec:demonstration}. Table \ref{tab:res:plain} contains the signature values for plain lowpass filter tests.
Table \ref{tab:res:ncc} contains the signatures from the extended \ac{NCC} trials.

\begin{table*}[!htbp]
\centering
\caption{Table of plain workflow signatures. Truncated from the original signature values to the first five characters for brevity. Signature standards are labeled as follows: RR - Rerun, RT - Repeat, RC - Recompute, RP - Reproduce, RPL-S - Replicate Scientifically, RPL-C - Replicate Computationally, RPL-T - Replicate Totally. Workflow trial names formatted as \emph{method-library-trial}.}
\label{tab:res:plain}
\begin{tabular}{@{}cccccccc@{}}
\toprule
Workflow Trial & RR    & RT    & RC    & RP    & RPL-S & RPL-C & RPL-T \\ \midrule
FFT-fftw-0     & 3cabf & deb85 & d5556 & 4fa43 & f9615 & f321d & ca359 \\
FFT-fftw-1     & 3cabf & deb85 & 117c4 & 9ae2d & bfc4b & 1e3cf & 4d937 \\
FFT-fftw-2     & 3cabf & deb85 & 03d3c & 26e2a & 8b785 & 17ca2 & 7fb60 \\
FFT-fftw-3     & 3cabf & deb85 & 3bc5e & 0d4a1 & 6648d & d0034 & 687bf \\
FFT-fftw-4     & 3cabf & deb85 & 58e17 & 13c75 & 33366 & c433b & 6d1fe \\
FFT-fftw-5     & 3cabf & deb85 & e1a4b & 2d0d5 & 4ce42 & 56002 & b4138 \\
FFT-fftw-6     & 3cabf & deb85 & 284d6 & b345f & 8c086 & fa69b & 589bb \\
FFT-fftw-7     & 3cabf & deb85 & a5f11 & dd944 & 5959c & 882e0 & 58b64 \\
FFT-fftw-8     & 3cabf & deb85 & c8e02 & 7b758 & 37c0d & 3f3c5 & 07808 \\
FFT-fftw-9     & 3cabf & deb85 & d9d6b & 0da65 & 7c3fb & 29eb0 & b6bcc \\
Pointwise-np-0 & 3cabf & 2f7f9 & 7cb68 & 3449a & 837f2 & 70370 & 7549c \\
Pointwise-np-1 & 3cabf & 2f7f9 & 080d7 & 928a7 & aa7a0 & 45b3a & 5b6a9 \\
Pointwise-np-2 & 3cabf & 2f7f9 & 24963 & 59f82 & 305bd & 36c5f & c7ce8 \\
Pointwise-np-3 & 3cabf & 2f7f9 & 6fa7a & efb06 & c2896 & ef68f & 2e278 \\
Pointwise-np-4 & 3cabf & 2f7f9 & 4b180 & b3742 & c0e64 & 5b7bb & 4eb40 \\
Pointwise-np-5 & 3cabf & 2f7f9 & 3e519 & f07ec & 65a85 & 0a35f & 00308 \\
Pointwise-np-6 & 3cabf & 2f7f9 & 0c776 & 501eb & b3fae & 8d837 & 3d6b0 \\
Pointwise-np-7 & 3cabf & 2f7f9 & 6d9bc & 6ebb8 & d4f24 & 60547 & f4bd6 \\
Pointwise-np-8 & 3cabf & 2f7f9 & 4019e & b9cf6 & fb869 & 2c487 & d09bc \\
Pointwise-np-9 & 3cabf & 2f7f9 & 91d11 & 5ce91 & 3be4d & b2ae1 & 9c36d \\
FFT-np-0       & 3cabf & dfbfa & 6d111 & bf3ac & c7cca & 3954f & 7f973 \\
FFT-np-1       & 3cabf & dfbfa & 9c1bc & c6a23 & dc639 & 0562e & e9632 \\
FFT-np-2       & 3cabf & dfbfa & 287a0 & 00598 & 8beb1 & 48677 & 6db23 \\
FFT-np-3       & 3cabf & dfbfa & dc330 & c52dc & 39eb8 & 62956 & d9c33 \\
FFT-np-4       & 3cabf & dfbfa & a33aa & 19172 & 819ce & 90cb3 & 4d5a6 \\
FFT-np-5       & 3cabf & dfbfa & 3d34d & 3880c & d23fe & 96fc3 & 337af \\
FFT-np-6       & 3cabf & dfbfa & 57c3b & 327c7 & a64b6 & a6687 & 3894f \\
FFT-np-7       & 3cabf & dfbfa & f2317 & d196f & 50607 & ba3ec & b47f6 \\
FFT-np-8       & 3cabf & dfbfa & 1732f & 2e491 & 8d5ab & 856a5 & eff00 \\
FFT-np-9       & 3cabf & dfbfa & 8ed58 & 5c0bf & abff8 & 8ed1b & bcc57 \\
FFT-cuda-0     & 3cabf & 1df3e & 4e149 & 77f90 & 4689c & 9bfa4 & 6eb83 \\
FFT-cuda-1     & 3cabf & 1df3e & 0fcc0 & f1f8c & 200d5 & 16eaa & b06e0 \\
FFT-cuda-2     & 3cabf & 1df3e & bddd5 & f9381 & 99da5 & b0cb3 & 23392 \\
FFT-cuda-3     & 3cabf & 1df3e & 8aae0 & f5a85 & 7320e & 6bc92 & 5deaa \\
FFT-cuda-4     & 3cabf & 1df3e & ac2fb & 0124a & 5d9b9 & 2ebe6 & 0f4c9 \\
FFT-cuda-5     & 3cabf & 1df3e & 6e775 & 55d9b & 261d5 & 375ca & f57c3 \\
FFT-cuda-6     & 3cabf & 1df3e & 9994d & d1cc2 & 6c0c1 & d5dcd & c497e \\
FFT-cuda-7     & 3cabf & 1df3e & 4c48b & 52427 & fd3f6 & 4ba05 & e4b81 \\
FFT-cuda-8     & 3cabf & 1df3e & 41009 & 150ef & fbbea & d777f & d71f9 \\
FFT-cuda-9     & 3cabf & 1df3e & eb6d0 & 9f45e & f7215 & ca7f8 & 55be9 \\ \bottomrule
\end{tabular}
\end{table*}

\begin{table*}[!htbp]
\centering
\caption{Table of \ac{NCC} workflow signatures. Truncated from the original signature values to the first five characters for brevity. Signature standards are labeled as follows: RR - Rerun, RT - Repeat, RC - Recompute, RP - Reproduce, RPL-S - Replicate Scientifically, RPL-C - Replicate Computationally, RPL-T - Replicate Totally. Workflow trial names formatted as \emph{method-library-trial}.}
\label{tab:res:ncc}
\begin{tabular}{@{}cccccccc@{}}
\toprule
Workflow Trial     & RR    & RT    & RC    & RP    & RPL-S & RPL-C & RPL-T \\ \midrule
Pointwise-np-ncc-0 & 8619d                  & c7529                  & 78d51                  & 8fd08                  & faf14                     & 998cf                     & f353b                     \\
Pointwise-np-ncc-1 & 8619d                  & c7529                  & 57d1d                  & bf29b                  & 52ae2                     & c7526                     & fcf65                     \\
Pointwise-np-ncc-2 & 8619d                  & c7529                  & 61f9a                  & bf29b                  & 52ae2                     & 95914                     & fcf65                     \\
Pointwise-np-ncc-3 & 8619d                  & c7529                  & 2d4d2                  & 8fd08                  & faf14                     & 2094b                     & f353b                     \\
Pointwise-np-ncc-4 & 8619d                  & c7529                  & 8128d                  & bf29b                  & 52ae2                     & 654a3                     & fcf65                     \\
Pointwise-np-ncc-5 & 8619d                  & c7529                  & c4615                  & bf29b                  & 52ae2                     & 52773                     & fcf65                     \\
Pointwise-np-ncc-6 & 8619d                  & c7529                  & 8772c                  & bf29b                  & 52ae2                     & 5d4fd                     & fcf65                     \\
Pointwise-np-ncc-7 & 8619d                  & c7529                  & 576e6                  & bf29b                  & 52ae2                     & faf9e                     & fcf65                     \\
Pointwise-np-ncc-8 & 8619d                  & c7529                  & 2e510                  & 8fd08                  & faf14                     & 0fc59                     & f353b                     \\
Pointwise-np-ncc-9 & 8619d                  & c7529                  & b5418                  & bf29b                  & 52ae2                     & 4b7b5                     & fcf65                     \\
FFT-fftw-ncc-0     & 8619d                  & 6bc99                  & 76aac                  & 8fd08                  & faf14                     & eac3f                     & 07d3a                     \\
FFT-fftw-ncc-1     & 8619d                  & 6bc99                  & 326f1                  & bf29b                  & 52ae2                     & 3748e                     & 9b208                     \\
FFT-fftw-ncc-2     & 8619d                  & 6bc99                  & c493b                  & bf29b                  & 52ae2                     & 64fa3                     & 9b208                     \\
FFT-fftw-ncc-3     & 8619d                  & 6bc99                  & 09956                  & 8fd08                  & faf14                     & dc375                     & 07d3a                     \\
FFT-fftw-ncc-4     & 8619d                  & 6bc99                  & e15ba                  & bf29b                  & 52ae2                     & 2a5df                     & 9b208                     \\
FFT-fftw-ncc-5     & 8619d                  & 6bc99                  & c261c                  & bf29b                  & 52ae2                     & ef2c5                     & 9b208                     \\
FFT-fftw-ncc-6     & 8619d                  & 6bc99                  & a3cc5                  & bf29b                  & 52ae2                     & a59c6                     & 9b208                     \\
FFT-fftw-ncc-7     & 8619d                  & 6bc99                  & 7295c                  & bf29b                  & 52ae2                     & 72836                     & 9b208                     \\
FFT-fftw-ncc-8     & 8619d                  & 6bc99                  & a03e0                  & 8fd08                  & faf14                     & 3131e                     & 07d3a                     \\
FFT-fftw-ncc-9     & 8619d                  & 6bc99                  & 88ff5                  & bf29b                  & 52ae2                     & 6d378                     & 9b208                     \\
FFT-np-ncc-0       & 8619d                  & b09c7                  & a6758                  & 8fd08                  & faf14                     & c6cb6                     & 89a7c                     \\
FFT-np-ncc-1       & 8619d                  & b09c7                  & d5ed5                  & bf29b                  & 52ae2                     & b21bb                     & cfb27                     \\
FFT-np-ncc-2       & 8619d                  & b09c7                  & 20176                  & bf29b                  & 52ae2                     & 84b38                     & cfb27                     \\
FFT-np-ncc-3       & 8619d                  & b09c7                  & ecc8b                  & 8fd08                  & faf14                     & 585e2                     & 89a7c                     \\
FFT-np-ncc-4       & 8619d                  & b09c7                  & 4a455                  & bf29b                  & 52ae2                     & 17d62                     & cfb27                     \\
FFT-np-ncc-5       & 8619d                  & b09c7                  & 10797                  & bf29b                  & 52ae2                     & 3119d                     & cfb27                     \\
FFT-np-ncc-6       & 8619d                  & b09c7                  & 147db                  & bf29b                  & 52ae2                     & 4ae6b                     & cfb27                     \\
FFT-np-ncc-7       & 8619d                  & b09c7                  & d181b                  & bf29b                  & 52ae2                     & 1962b                     & cfb27                     \\
FFT-np-ncc-8       & 8619d                  & b09c7                  & 3bc4c                  & 8fd08                  & faf14                     & 2f68d                     & 89a7c                     \\
FFT-np-ncc-9       & 8619d                  & b09c7                  & e9262                  & bf29b                  & 52ae2                     & 9518f                     & cfb27                     \\
FFT-cuda-ncc-0     & 8619d                  & 85f98                  & 09b82                  & 8fd08                  & faf14                     & 9343a                     & 7ecb0                     \\
FFT-cuda-ncc-1     & 8619d                  & 85f98                  & 0ebd4                  & bf29b                  & 52ae2                     & 2df4d                     & 6f8a9                     \\
FFT-cuda-ncc-2     & 8619d                  & 85f98                  & 641dc                  & bf29b                  & 52ae2                     & 78071                     & 6f8a9                     \\
FFT-cuda-ncc-3     & 8619d                  & 85f98                  & 7e4bf                  & 8fd08                  & faf14                     & d7105                     & 7ecb0                     \\
FFT-cuda-ncc-4     & 8619d                  & 85f98                  & 59a07                  & bf29b                  & 52ae2                     & 4f18d                     & 6f8a9                     \\
FFT-cuda-ncc-5     & 8619d                  & 85f98                  & fb74a                  & bf29b                  & 52ae2                     & 46996                     & 6f8a9                     \\
FFT-cuda-ncc-6     & 8619d                  & 85f98                  & 6b8ea                  & bf29b                  & 52ae2                     & 25474                     & 6f8a9                     \\
FFT-cuda-ncc-7     & 8619d                  & 85f98                  & 8581f                  & bf29b                  & 52ae2                     & a9abf                     & 6f8a9                     \\
FFT-cuda-ncc-8     & 8619d                  & 85f98                  & 900e1                  & bf29b                  & 52ae2                     & 93555                     & 6f8a9                     \\
FFT-cuda-ncc-9     & 8619d                  & 85f98                  & 024b7                  & bf29b                  & 52ae2                     & 1dfe3                     & 6f8a9                     \\ \bottomrule
\end{tabular}
\end{table*}

\clearpage
\bibliographystyle{elsarticle-num} 
\bibliography{bibliography.bib}

\hfill
\setlength\intextsep{0pt} 
\begin{wrapfigure}{l}{0.13\textwidth}
    \centering
    \includegraphics[width=0.15\textwidth]{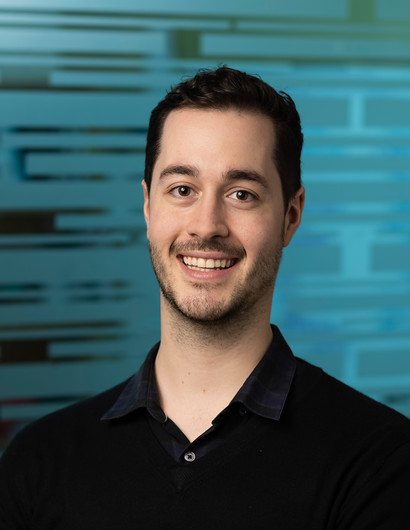}
\end{wrapfigure}
\noindent \textbf{Nicholas J. Pritchard} received the BSc. in computer science and software engineering in 2018 and the MPhil in Physics in 2021 from the University of Western Australia, Perth, Australia.
He is currently pursuing a Ph.D. in physics at the University of Western Australia, Perth, Australia.
From 2021 to 2023, he was a Research Assistant and Software Engineer with the International Centre for Radio Astronomy Research, University of Western Australia, Perth, Australia.
His research interests include using neuromorphic computing, spiking neural networks, and data-intensive workflow systems.
\subsection*{ }
\setlength\intextsep{0pt} 
\begin{wrapfigure}{l}{0.13\textwidth}
    \centering
    \includegraphics[width=0.15\textwidth]{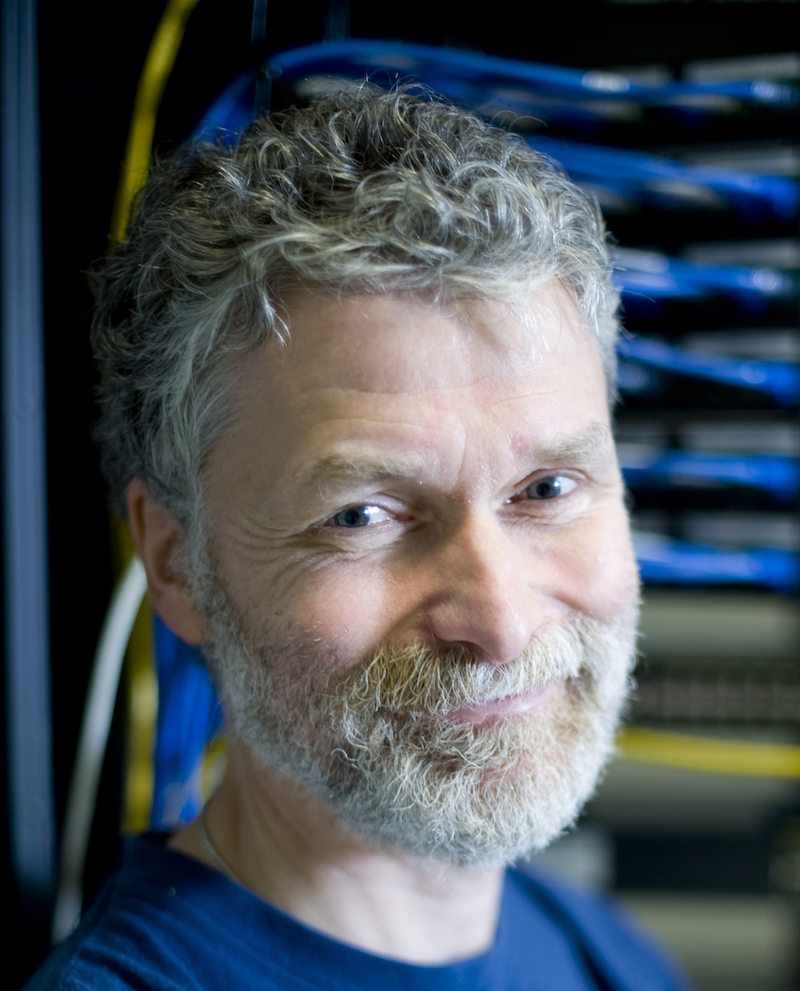}
\end{wrapfigure}
\noindent \textbf{Andreas Wicenec} is Professor at the University of Western Australia since 2010, leading the Data Intensive Astronomy Program of the International Centre for Radio Astronomy Research.
He joined ESO in 1997 as an archive specialist and was involved in the final implementation of the archive for ESO's Very Large Telescope (VLT) and then became ESO's Archive Scientist and led the ALMA archive subsystem development group. 
Prof. Wicenec's scientific interests and publications include high-precision global astrometry, optical background radiation, stellar photometry, planetary nebulae dynamics and evolution, observational survey astronomy, and related scheduling and computational concepts.
\end{document}